\documentclass[aps,prx,twocolumn,amsfonts,amssymb,amsmath,showpacs,
floatfix,nofootinbib,groupedaddress,superscriptaddress,citesort]{revtex4}
\usepackage{mathrsfs}
\usepackage{amsfonts}
\usepackage{amstext}
\usepackage{amsmath}
\usepackage{amssymb}
\usepackage{bm}% bold math
\usepackage{CJK}
\usepackage{bbm}
\usepackage[dvips]{graphicx}
\def\qed{\leavevmode\unskip\penalty9999 \hbox{}\nobreak\hfill
     \quad\hbox{\leavevmode  \hbox to.77778em{%
              \hfil\vrule   \vbox to.675em%
               {\hrule width.6em\vfil\hrule}\vrule\hfil}}
     \par\vskip3pt}

\usepackage{amssymb}
\usepackage{graphicx}
\usepackage{graphics}
\usepackage{amsmath}
\usepackage{amsthm}
\usepackage{color}
\usepackage{dsfont}
\definecolor{darkred}  {rgb}{0.5,0,0}
\definecolor{darkblue} {rgb}{0,0,0.5}
\definecolor{darkgreen}{rgb}{0,0.5,0}
\usepackage{hyperref}
\hypersetup{
	pdftitle = {QRT Proposal},
	pdfauthor = {},
	colorlinks = true,
	urlcolor  = blue,         % color of external links
	linkcolor = red,     % color of internal links
	citecolor = blue,    % color of links to bibliography
	filecolor = darkred       % color of file links
}
\usepackage{mathtools}
\def\ra{\rangle}
\def\la{\langle}

\def\ot{\otimes}
\newtheorem{theorem}{Theorem}

\newtheorem{lemma}[theorem]{Lemma}

\newtheorem{pro}[theorem]{Proposition}

\newcommand{\bea}{\begin{eqnarray}}
\newcommand{\eea}{\end{eqnarray}}
\newcommand{\be}{\begin{equation}}
\newcommand{\ee}{\end{equation}}
\newcommand{\ba}{\begin{equation}\begin{aligned}}
\newcommand{\ea}{\end{aligned}\end{equation}}
\newcommand{\rank}{\text{Rank}}

\newcommand{\beax}{\begin{eqnarray*}}
\newcommand{\eeax}{\end{eqnarray*}}
\newcommand{\bex}{\begin{equation*}}
\newcommand{\eex}{\end{equation*}}

\newtheorem{definition}{Definition}
\theoremstyle{remark}
\newtheorem{remark}{Remark}

\def\be{\begin{equation}}
\def\ee{\end{equation}}

\newcommand{\mH}{\mathcal{H}}

\newcommand{\mK}{\mathcal{K}}

\newcommand{\mP}{\mathcal{P}}

\newcommand{\mS}{\mathcal{S}}

\newcommand{\lr}{\rangle\langle}

\newcommand{\tr}{{\rm Tr}}

\newcommand{\mbb}[1]{\mathbb{#1}}

\newcommand{\mbR}{\mathbb{R}}

\def\>{\rangle}
\def\<{\langle}

\begin{document}

%========================================================================================%

\preprint{APS/123-QED}
\begin{CJK*}{GB}{gbsn}
\title{Multipartite Entanglement Measure and Complete Monogamy Relation\\}% Force line breaks with \\

%========================================================================================%

\author{Yu Guo}
\email{guoyu3@aliyun.com}
\affiliation{Institute of Quantum Information Science, School of Mathematics and Statistics, Shanxi Datong University, Datong, Shanxi 037009, China}

%========================================================================================%

\author{Lin Zhang}
\email{godyalin@163.com} 
\affiliation{Institute
of Mathematics, Hangzhou Dianzi University, Hangzhou 310018,
PR~China} 
\affiliation{Max-Planck-Institute for Mathematics in the
Sciences, Leipzig 04103, Germany}

%========================================================================================%

\begin{abstract}
	Although many different entanglement measures have been proposed so far,
	much less is known in the multipartite case, which leads to the previous 
	monogamy relations in literatures are not complete.
	We establish here a strict framework for defining multipartite entanglement
	measure (MEM): apart from the postulates of bipartite measure 
	[i.e., vanishing on separable and nonincreasing under local operations and classical communication (LOCC)], 
	a genuine MEM should additionally satisfy the \textit{unification condition} 
	and the \textit{hierarchy condition}. We then come up with a \textit{complete monogamy} 
	formula for the unified MEM (an MEM is called a unified MEM if it satisfies the unification condition)
	and a {\it tightly complete monogamy} relation for the genuine MEM (an MEM is 
	called a genuine MEM if it satisfies both the unification condition and the 
	hierarchy condition). Consequently, we propose MEMs which are multipartite 
	extensions of entanglement of formation (EoF), concurrence, tangle, Tsallis 
	$q$-entropy of entanglement, R\'{e}nyi $\alpha$-entropy of entanglement, 
	the convex-roof extension of negativity and negativity, respectively. We show that   
	(i) the extensions of EoF, concurrence, tangle, and Tsallis $q$-entropy of entanglement
	are genuine MEMs, (ii) multipartite extensions of R\'{e}nyi $\alpha$-entropy of 
	entanglement, negativity and the convex-roof extension of negativity are unified 
	MEMs but not genuine MEMs, and (iii) all these multipartite extensions are 
	completely monogamous and the ones which are defined by the convex-roof structure 
	(except for the R\'{e}nyi $\alpha$-entropy of entanglement and the convex-roof 
	extension of negativity) are not only completely monogamous but also tightly 
	completely monogamous. In addition, as a by-product, we find out a class of 
	states that satisfy the additivity of EoF. We also find a class of tripartite 
	states that one part can maximally entangled with other two parts simultaneously 
	according to the definition of maximally entangled mixed state (MEMS) in 
	[Quantum Inf. Comput. 12, 0063 (2012)]. Consequently, we improve the definition 
	of maximally entangled state (MES) and prove that there is no MEMS and that 
	the only MES is the pure MES.
		
	%\begin{description}
	%\item[PACS numbers]03.67.Mn, 03.65.Db, 03.65.Ud.
	%\end{description}
\end{abstract}

%========================================================================================%

\maketitle
\end{CJK*}
%\tableofcontents

%========================================================================================%

\section{Introduction}

Entanglement is recognized as the most
important resource in quantum information processing tasks~\cite{Nielsen}. A
fundamental problem in this field is to quantify entanglement.
Many entanglement measures have been proposed for this
purpose, such as the distillable entanglement~\cite{Bennett1996pra},
entanglement cost~\cite{Bennett1996pra,Bennett1996prl}, 
entanglement of formation~\cite{Bennett1996prl,Horodecki01},
concurrence~\cite{Hill,Wootters,Rungta}, tangle~\cite{Rungta2003pra},
relative entropy of entanglement~\cite{Vedral97,Vedral98}, negativity~\cite{Vidal02,Lee},
geometric measure~\cite{Shimony95,Wei2003pra,Barnum}, squashed
entanglement~\cite{Christandl2004jmp,Yang2009ieee}, the
conditional entanglement of mutual information~\cite{Yang2008prl},
three-tangle~\cite{Coffman}, the generalizations of  concurrence~\cite{Hong2012pra,Hiesmayr2008pra}, 
and the $\alpha$-entanglement entropy~\cite{Szalay}, etc. However, 
apart from the $\alpha$-entanglement entropy, all other measures are either only defined on the
bipartite case or just discussed with only the axioms of the bipartite case.

One of the most important issues closely related to entanglement measure is the
monogamy relation of entanglement~\cite{Terhal2004}, which states
that, unlike classical correlations, if two parties $A$ and $B$ are
maximally entangled, then neither of them can share entanglement
with a third party $C$. Entanglement monogamy has many applications not only in quantum physics~\cite{Bennett2014,Toner,Seevinck} but also
in other area of physics, such as no-signaling theories~\cite{streltsov2012are}, 
condensed matter physics~\cite{Ma2011,Brandao2013,Garcia}, 
statistical mechanics~\cite{Bennett2014}, and even black-hole physics~\cite{Lloyd}.
Particularly, it is the crucial property 
that guarantees quantum key distribution secure~\cite{Terhal2004,Pawlowski}.
An important basic issue in this field is to
determine whether a given entanglement measure is monogamous.
Considerable efforts have been devoted to this task in the last two
decades~\cite{Coffman,Zhuxuena2014pra,Osborne,streltsov2012are,Lan16,
Ouyongcheng2007pra2,Kim2009,kim2012limitations,Kumar,Deng,Karczewski,
Bai,Oliveira2014pra,Koashi,
Luo2016pra,Dhar,Chengshuming,Allen,Hehuan,GG2019,GG,G2019,Camalet}
ever since Coffman, Kundu, and Wootters (CKW) presented the first
quantitative monogamy relation in Ref.~\cite{Coffman} for
three-qubit states. So far, we have known that the one-way
distillable entanglement~\cite[Theorem 6]{Koashi} and squashed
entanglement~\cite[Theorem 8]{Koashi} and all the other measures
that defined by the convex-roof extension are
monogamous~\cite{GG2019}. But all these monogamy relations are
discussed via the bipartite measures of entanglement: only the
relation between $A|BC$, $AB$ and $AC$ are revealed, the global correlation
in ABC and the correlation contained in part $BC$ is missed~[see
Eqs.~\eqref{definition-bipartite-monogamy} and~\eqref{power1}
below],  where the vertical bar indicates 
the bipartite split across which we will measure the (bipartite) 
entanglement. From this point of view, the monogamy relation in the sense
of CKW is not ``complete''. We thus need to explore a complete monogamy
relation which can exhibit the entanglement between $ABC$, $AB$,
$AC$ and $BC$ in extenso.

The phenomenon becomes much more complex when moving from the
bipartite case to the multipartite case~\cite{Horodecki2009,Gao2010pra,Gao2014prl,Szalay}.
For an $m$-partite system, we have to encounter entanglement for both $m$-partite and 
$k$-partite cases, $k\leqslant m$. Particularly, a ``complete monogamy relation'' 
involves both MEM and bipartite ones, which requires a ``unified'' way (i.e., the unification condition) 
to define entanglement measures. In~\cite{Szalay}, Szalay developed the two kinds of indicator functions for characterizing mulitpartite entanglement based on the complex lattice-theoretic structure of 
partial separability classification for multipartite states. But the second kind in fact can 
not quantify entanglement effectively and the unification condition 
was not considered as a necessary requirement of MEM. The purpose of this paper is to give, 
concisely, ``richer'' postulates in defining a genuine MEM
from which we can quantify and compare the amount of entanglement for 
both bipartite and multipartite systems in a unified way. 
We then explore the complete monogamy relation under these postulates and 
illustrate with several MEMs which are multipartite extensions of EoF, concurrence, tangle,
Tsallis $q$-entropy of entanglement, R\'{e}nyi $\alpha$-entropy of entanglement, negativity, 
and the convex-roof extension of negativity. Hereafter, we let $\mH^{ABC}$
be a tripartite Hilbert space with finite dimension and let $\mS^{X}$ be the set of density
operators acting on $\mH^{X}$.

The rest of this paper is organized as follows. We review the postulates of bipartite entanglement measure
and the associated monogamy relation in Sec. II, and explore the additional postulates for multipartite
entanglement measures in Sec. III. Sec. IV proposes the complete monogamy relation and the tight complete monogamy relation for multipartite measures with the additional postulates. We then extend some well-known bipartite entanglement measures to tripartite case, and discuss their complete monogamy property. Particularly, we find a class of states that are additive under the tripartite entanglement of formation.
Sec. VI mainly discusses what is the maximally entangled state. We give a new definition of maximally entangled 
state by means of its extension. Finally, in Sec. VII, we summarize our main findings and conclusions.

\section{Reviewing of the bipartite entanglement measure}

We begin with reviewing the
bipartite entanglement measure. A function $E:
\mS^{AB}\to\mbb{R}_{+}$ is called an \emph{entanglement measure} if
it satisfies~\cite{Vedral97}: 
\begin{itemize}
	\item (E-1) $E(\rho)=0$ if $\rho$ is
	separable; 
	\item (E-2) $E$ cannot increase under LOCC, i.e.,
	$E(\Phi(\rho))\leqslant E(\rho)$ for any LOCC $\Phi$ [(E-2) 
	implies that $E$ is invariant under local unitary
	operations, i.e., $E(\rho)=E(U^A\otimes U^B\rho U^{A,\dag}\otimes
	U^{B,\dag})$ for any local unitaries $U^A$ and $U^B$].
	The map $\Phi$ is completely positive and trace preserving 
	(CPTP). 
\end{itemize}
In general, LOCC can be stochastic, 
in the sense that $\rho$ can be converted to $\sigma_{j}$ with some probability $p_j$. 
(It is possible that $E(\sigma_{j_0})>E(\rho)$ for some $j_0$.)
In this case, the map from $\rho$ to $\sigma_{j}$ can not be described in general by a CPTP map. 
However, by introducing a ``flag'' system $A'$, we can view the ensemble $\{\sigma_{j},p_j\}$ 
as a classical quantum state $\sigma':=\sum_{j}p_j|j\lr j|^{A'}\otimes\sigma_{j}$. 
Hence, if $\rho$ can be converted by LOCC to $\sigma_{j}$ with probability $p_j$, 
then there exists a CPTP LOCC map $\Phi$ such that $\Phi(\rho)=\sigma'$.
Therefore, the definition above of a measure of entanglement captures also probabilistic transformations. 
Particularly, $E$ must satisfy $E\left(\sigma'\right)\leq E\left(\rho\right)$.

Almost all measures of entanglement studied in literature (although not all~\cite{Plenio2005}) satisfy
\be\label{average}
E\left(\sigma'\right)=\sum_{j}p_jE(\sigma_{j})\;,
\ee
which is very intuitive since $A'$ is just a classical system encoding the value of $j$. 
In this case the condition $E\left(\sigma'\right)\leq E\left(\rho\right)$ 
becomes 
\beax
\sum_{j}p_jE(\sigma_{j})\leq E\left(\rho\right).
\eeax That is, LOCC can 
not increase entanglement on average. 
An entanglement measure $E$ is said to be an entanglement
monotone~\cite{Vidal2000} if it satisfies Eq.~\eqref{average} and is convex additionally.

Let $E$ be a bipartite measure of entanglement.
The entanglement of formation associated with $E$,
denoted by $E_F$, is defined as the average pure-state measure minimized
over all pure-state decompositions
\begin{align}\label{bipartite-eofmin}
E_F\left(\rho\right):=\min\sum_{j=1}^{n}p_jE\left(|\psi_j\lr\psi_j|\right),
\end{align}
which is also called the convex-roof extension of $E$.
In general, for
pure state $|\psi\rangle\in\mH^{AB}$, $\rho^A={\rm
Tr}_B|\psi\rangle\langle\psi|$, 
\be\label{h}
E\left( |\psi\lr\psi|\right)= h\left(
\rho^A\right)
\ee
for some positive function $h$.
Vidal~\cite[Theorem 2]{Vidal2000}
showed that $E_F$, defined as Eqs.~\eqref{bipartite-eofmin} and~\eqref{h}, 
is an entanglement monotone iff $h$ is also
\emph{concave}, i.e.
\be \label{concave}
h[\lambda\rho_1+(1-\lambda)\rho_2]\geqslant\lambda
h(\rho_1)+(1-\lambda)h(\rho_2)
\ee
for any states $\rho_1$,
$\rho_2$, and any $\lambda\in[0,1]$.
Very recently, Guo and Gour~\cite{GG2019} showed that, if $h$ is strictly concave,
then $E_F$ is monogamous, i.e., for any $\rho^{ABC}\in\mS^{ABC}$ 
that satisfies the disentangling condition
\bea\label{definition-bipartite-monogamy}
E_F(\rho^{AB})=E_F(\rho^{A|BC})
\eea
we have that $E_F(\rho^{AC})=0$,
or equivalently (for continuous measures~\cite{GG}), 
there exists some $\alpha>0$ such that
\bea\label{power1}
E_F^\alpha(\rho^{A|BC})\geqslant E_F^\alpha(\rho^{AB})+ E_F^\alpha(\rho^{AC})
\eea
holds for all $\rho^{ABC}\in\mS^{ABC}$.

For convenience, we list some bipartite entanglement measures.
The first convex-roof extended measure is entanglement of formation 
(EoF)~\cite{Bennett1996pra,Horodecki01}, $E_f$,
 which is defined by
 \be
 E_f(|\psi\ra)=E(|\psi\ra):=S(\rho^A),\quad \rho^A=\tr_B|\psi\ra\la\psi|,
 \ee
 for pure state $|\psi\ra\in\mH^{AB}$, where $S(\rho):=-\tr(\rho\ln\rho)$ is the von Neumann entropy,
 and
 \bea
 E_f(\rho):=\min_{\{p_i,|\psi_i\ra\}}\sum_ip_iE(|\psi_i\ra)
 \eea
 for mixed state, where the minimum is taken over all
 pure-state decomposition $\{p_i,|\psi_i\ra\}$ of $\rho\in\mS^{AB}$ 
 (Throughout this paper, we identify the
 original bipartite entanglement of formation with $E_f$,
 the notation $E_F$ with capital $F$ in the subscription 
 denotes other general convex-roof extended measures).
For bipartite pure state $|\psi\ra\in\mH^{AB}$,
concurrence~\cite{Hill,Wootters,Rungta} and tangle~\cite{Rungta2003pra} are defined by
\beax 
C(|\psi\ra)=\sqrt{2[1-{\rm Tr}( \rho^A) ^2}]
\eeax
and \beax
\tau(|\psi\ra)=C^2(|\psi\ra),
\eeax
respectively.
For mixed state, they are defined by the convex-roof extension as Eq.~\eqref{bipartite-eofmin}.
The negativity~\cite{Vidal02,Lee} is defined by
\beax 
N(\rho)=\frac{1}{2}( \|\rho^{T_a}\|_\tr-1),\quad \rho\in\mS^{AB},
\eeax
where $T_x$ 
denotes the transpose with respect to
the subsystem $X$, $\|\cdot\|_{\tr}$ denotes the trace norm.
The convex-roof extension of $N$, $N_F$ is 
defined as Eq.~\eqref{bipartite-eofmin} (i.e., taking $E=N$).
Any function that can be expressed as
\be
H_g(\rho)=\tr[g(\rho)]=\sum_jg(p_j)\,,
\ee
where $p_j$s are the eigenvalues of $\rho$ is strictly concave if $g''(p)<0$ for all $0<p<1$~\cite{GG2019}. 
This includes the quantum Tsallis $q$-entropy~\cite{Tsallis,Landsberg,Raggio} $T_q$
with $q>0$ and the R\'enyi $\alpha$-entropy~\cite{Greenberger,Renyi,Dur} $R_\alpha$ with
$\alpha\in[0,1]$.
Consequently, according to Eq.~\eqref{definition-bipartite-monogamy}, 
it is proved that all bipartite entanglement monotones are monogamous
for pure states and all $E_F$ in the literatures so far, such as $E_f$, $C$, $\tau$, $N_F$,
Tsallis $q$-entropy of entanglement ($q>0$) and
R\'{e}nyi $\alpha-$ entropy of entanglement ($0<\alpha< 1$), 
etc., are monogamous~\cite{GG2019}.

\section{Postulates for multipartite entanglement measure}

\subsection{Multipartite entanglement monotone}

We now turn to discussion of multipartite measures of entanglement. 
A function 
$E^{(m)}: \mS^{A_1A_2\cdots A_m}\to\mbb{R}_{+}$ 
is called a $m$-partite entanglement measure in 
literatures~\cite{Horodecki2009,Hong2012pra,Hiesmayr2008pra} if
it satisfies:
\begin{itemize}
	\item {\bf(E1)} $E^{(m)}(\rho)=0$ if $\rho$ is fully separable;
	\item {\bf(E2)}
	$E^{(m)}$ cannot increase
	under $m$-partite LOCC.
\end{itemize}
 In addition, $E^{(m)}$ is said to be 
an $m$-partite entanglement monotone if it is convex and 
does not increase on average under $m$-partite stochastic LOCC.
For simplicity,
throughout this paper,
we call $E_F^{(m)}$ defined as
\bea\label{m-partite-eofmin}
E_F^{(m)}(\rho):=\min\sum_ip_iE^{(m)}(|\psi_i\ra)
\eea
an $m$-partite entanglement of formation associated with
$E^{(m)}$ provided that $E^{(m)}$ is an $m$-partite 
entanglement measure on pure states. From now on,
we only consider the tripartite system $\mH^{ABC}$ unless otherwise stated,
and the case for $m\geqslant3$ could be argued analogously.
As a generalization of Vidal's scenario for bipartite entanglement 
monotone proposed in Ref.~\cite{Vidal2000}, we give at first a 
necessary-sufficient criterion of tripartite entanglement monotone (TEM):

\begin{pro}\label{m-monotone}
	Let $E^{(3)}: \mH^{ABC}\rightarrow\mbR_+$ be a function that defined by
	\bea\label{m-h}
	E^{(3)}(|\psi\ra)=h^{(3)}(\rho^A\otimes\rho^B\otimes\rho^C),\quad |\psi\ra\in\mH^{ABC}.
	\eea
	and let $E_F^{(3)}$ be a function defined as Eq.~\eqref{m-partite-eofmin}.
	Then $E_F^{(3)}$ is a TEM if and only if
	(i) $h^{(3)}$ is invariant under local unitary operations, and (ii) $h^{(3)}$ is LOCC-concave,
	i.e.,
	\bea\label{locc-concave}
	h^{(3)}\left( \rho^A\otimes\rho^B\otimes\rho^C\right) \geqslant \sum_kp_kh^{(3)}\left(   \sigma_k^A\otimes\sigma_k^B\otimes\sigma_k^C  \right)
	\eea
	holds for any stochastic LOCC $\{\Phi_k\}$ acting on $|\psi\ra\la\psi|$,
	where $\sigma_k^{x}=\tr_{\bar{x}}\sigma_k$, $p_k\sigma_k=\Phi_k(|\psi\ra\la\psi|)$.
\end{pro}

\begin{proof}
	According to the scenario in Ref.~\cite{Vidal02},
	we only need to consider a family $\{\Phi_k\}$ consisting of 
	completely positive linear maps
	such that $\Phi_k(\rho)=p_k{\sigma_k}$,
	where 
	\beax
	\Phi_k(\rho)=M_k\rho M_k^\dag=M^A_k\otimes I^{BC}\rho M^{A,\dag}_k\otimes I^{BC}
	\eeax 
	transforms pure states to some scalar multiple of pure states,
	$\sum_kM^{A,\dag}_k M^A_k= I^A$.
	We assume at first that the initial state $\rho\in\mS^{ABC}$ is pure.
	Then it yields that $E^{(3)}(\rho)\geqslant\sum_kp_kE^{(3)}(\sigma_k)$ 
	holds iff $h^{(3)}$ is LOCC-concave.
	Apparently, $E^{(3)}(\rho)=h^{(3)}\left(
	\rho^A\otimes\rho^B\otimes\rho^C\right)$ and $E^{(3)}(\sigma_k)=h^{(3)}\left(
	\sigma^A_k\otimes\sigma^B_k\otimes\sigma^C_k\right)$ since $\sigma_k$ 
	still is a pure state for each $k$. Therefore,
	the inequality $E^{(3)}(\rho)\geqslant\sum_kp_kE^{(3)}(\sigma_k)$ 
	can be rewritten as
	\begin{eqnarray*}
		h^{(3)}\left( \rho^A\otimes\rho^B\otimes\rho^C\right)
		\geqslant\sum_{k}p_kh^{(3)}(\sigma_k^A\otimes\sigma^B_k\otimes\sigma^C_k).
	\end{eqnarray*}
	That is, if $h^{(3)}$ is LOCC-concave, then $E^{(3)}$ does not 
	increase on average under LOCC for pure states and vice versa.
	So it remains to show that $E_F^{(3)}$ does not increase on 
	average under LOCC for mixed states
	with the assumption that $h^{(3)}$ is LOCC-concave.
	For any mixed state $\rho\in\mS^{ABC}$, there exists an ensemble
	$\{t_j,|\eta_j\rangle\}$
	such that
	\beax
	E_F^{(3)}(\rho)=\sum_jt_jE^{(3)}(|\eta_j\rangle).
	\eeax
	For each $j$, let
	\begin{eqnarray*}
		t_{jk}\sigma_{jk}=\Phi_{k}(|\eta_j\rangle\langle\eta_j|),\quad t_{jk}={\rm
			Tr}[\Phi_{k}(|\eta_j\rangle\langle\eta_j|)].
	\end{eqnarray*}
	Then we achieve that
	\begin{eqnarray*}
		E_F^{(3)}(\rho)&=&\sum_jt_jE^{(3)}(|\eta_j\rangle)
		\geqslant
		\sum_{j,k}t_jt_{jk}E^{(3)}(\sigma_{jk})\\
		&\geqslant&\sum_kp_kE_F^{(3)}(\sigma_k),
	\end{eqnarray*}
	where ${p_k}=\sum_jt_jt_{jk}$
	In addition, it is well-known that entanglement is 
	invariant under local unitary operation, which is equivalent to
	the fact that $h$ is invariant under local unitary operation.
	The proof is completed.
\end{proof}

\begin{remark}
	The inequality~\eqref{locc-concave} in Condition ii) above reduces to 
	Eq.~\eqref{concave} for bipartite case.
	That is, for bipartite case, concavity is equivalent to
	LOCC-concavity, but it is unknown whether it also true for tripartite case.
\end{remark}

\subsection{Unification condition for multipartite entanglement measure}

As mentioned before, for MEM, a natural question that arisen from the monogamy 
relation is whether it obeys:
\begin{itemize}
	\item {\bf(E3)}:~\textit{the unification condition}, i.e., 
	$E^{(3)}$ is consistent with $E^{(2)}$.
\end{itemize}
That is, when we analyze the entanglement contained in a given tripartite state $\rho^{ABC}\in\mS^{ABC}$,
we have to couple with not only the total entanglement in $\rho^{ABC}$ measured by $E^{(3)}$ but also
the entanglement in $\rho^{AB}$, $\rho^{AC}$, $\rho^{BC}$, $\rho^{A|BC}$, $\rho^{B|AC}$, and $\rho^{AB|C}$
measured by $E^{(2)}$, and thus $E^{(3)}$ and $E^{(2)}$ must be defined in the same way.
Then, how can we define them in the same way?
We begin with a simple observation.
Let $|\psi\ra^{ABC}$ be a bi-separable pure state in $\mH^{ABC}$, e.g., 
$|\psi\ra^{ABC}=|\psi\ra^{AB}|\psi\ra^{C}$.
It is clear that, the only entanglement of such a state is contained in $|\psi\ra^{AB}$,
namely, we must have 
\bea\label{unification1}
E^{(3)}(|\psi\ra^{ABC})=E^{(2)}(|\psi\ra^{AB}).
\eea
In this way, we can find the link between
 $E^{(2)}$ and $E^{(3)}$ (or $h^{(2)}$ and $h^{(3)}$). 
For instance, if
$E^{(3)}(|\psi\ra^{ABC})=h^{(3)}(\rho^A\otimes\rho^B\otimes\rho^C)$,
 we have $E^{(2)}(|\psi\ra^{AB})=h^{(2)}(\rho^A\otimes\rho^B)$ with
the same ``action'' of function $h$ [e.g., EoF and the tripartite EoF (also see in Sec. V): $E^{(2)}(|\psi\ra^{AB})=h^{(2)}(\rho^A\otimes\rho^B)=\frac12S(\rho^A\otimes\rho^B)$ while $E^{(3)}(|\psi\ra^{ABC})=h^{(3)}(\rho^A\otimes\rho^B\otimes\rho^C)=\frac12S(\rho^A\otimes\rho^B\otimes\rho^C)$]. 
In general, 
$E^{(2)}$ is uniquely determined by
$E^{(3)}$ but not vice versa.
It is worth mentioning that,
$h^{(2)}(\rho^A\otimes\rho^B)$ can be instead by $h(\rho^A)$ since any
bipartite pure state has Schmidt decomposition, which guarantees
that the eigenvalues of $\rho^A$ coincide with that of $\rho^B$.
That is, $h(\rho^A)$ is in fact $h^{(2)}(\rho^A\otimes\rho^B)$, and part
$A$ and part $B$ are symmetric, or equivalently, 
\beax 
h^{(2)}(\rho^A\otimes\rho^B)=h^{(2)}(\rho^B\otimes\rho^A).
\eeax 
So, as one may
expect, for multiparite case, the unification condition requires the
measure of multipartite entanglement must be 
\textit{invariant under the permutations of the subsystems}. 
Namely, the amount of entanglement contained in a state is fixed: 
\bea\label{unification2}
E^{(3)}(\rho^{ABC})=E^{(3)}(\rho^{\pi(ABC)}),
\eea 
where $\pi$ is a permutation of the subsystems 
[note that $E(\rho^{A|BC})\neq E(\rho^{X|YZ})$ 
in general whenever $X\neq A$, $X,Y,Z\in\{A,B,C\}$]. 
In addition, we always have
\bea\label{unification3}
E^{(3)}(ABC)\geqslant E^{(2)}(XY),\quad X,Y,\in\{A,B,C\}
\eea
 since the partial trace is a special
LOCC. $E^{(3)}$ is called a \textit{unified}
multipartite entanglement measure if it satisfies (E3). Hereafter,
we always assume that $E^{(3)}$ is a unified measure unless
otherwise specified.

We need note here that, although the analytic formulas for $E^{(2)}$ and
$E^{(3)}$ can not be uniquely determined each other, namely, the ``same action''
of $h$ has a little ambiguity since they are defined on different systems,
$E^{(2)}$ can be uniquely determined for any given
$E^{(3)}$ by the requirements in Eqs.~\eqref{unification1} and~\eqref{unification2}
generally.

\subsection{Hierarchy condition for multipartite entanglement measure}

There are different kinds of separability in
the tripartite case: fully separable state, $2$-partite separable state and
genuinely entangled state. We denote by $E^{(3-2)}$ the
$2$-partite entanglement measure associated with $E^{(3)}$, which is defined
by
\bea E^{(3-2)}(|\psi\ra)&:=&\min\{E^{(2)}(|\psi\ra^{A|BC}),
E^{(2)}(|\psi\ra^{AB|C}),\nonumber\\
&&~~~~~~E^{(2)}(|\psi\ra^{B|AC})\}.
\eea
For any given $\rho^{ABC}\in\mS^{ABC}$, $E^{(3)}(\rho^{ABC})$
extract the ``total entanglement'' contained in the state while
$E^{(2)}(\rho^{X|YZ})$ only quantifies the ``bipartite
entanglement'' up to some bipartite cutting $X|YZ$,
$X,Y,Z\in\{A,B,C\}$. For instance, for any entanglement monotone $E$, the pure state $|\psi\ra^{ABC}$
that satisfying the disentangling condition $E(|\psi\ra^{A|BC})=E(\rho^{AB})$ has the form
of $|\psi\ra^{AB_1}|\psi\ra^{B_2C}$ for some subspace $\mH^{B_1B_2}$ in $\mH^B$~\cite{Hehuan,GG,GG2019}.
In such a case, $E(A|BC)$ only reflects the entanglement between $A$ and $BC$, 
the entanglement between $B$ and $C$ is missed whenever $|\psi\ra^{B_2C}$ 
is entangled (in fact, $|\psi\ra^{B_2C}$ can be a maximally entangled sate, also see in Sec.~VI). We thus need additionally the following
\textit{hierarchy condition}:
\begin{itemize}
	\item {\bf
		(E4)}:~$E^{(3)}(\rho^{ABC})\geqslant
	E^{(2)}(\rho^{X|YZ})\geqslant E^{(3-2)}(\rho^{ABC})$
	holds for all $\rho^{ABC}$, $X,Y,Z\in\{A,B,C\}$.
\end{itemize}
 That is, a nonnegative function $E^{(3)}$, as
a ``genuine'' tripartite entanglement measure, not only need
obey the conditions (E1)-(E2) but also need satisfy the conditions (E3)
and (E4).
One can easily check that the triparite squashed entanglement
and the tripartite conditional entanglement of mutual information 
are genuine entanglement monotones
[i.e., they also satisfy (E3)-(E4)], but the $k$-ME concurrence~\cite{Hong2012pra} 
violates (E4), and the three-tangle is even
not a unified measure
(note that the three-tangle, denoted by $\tau_{ABC}$,
is defined by
\beax
\tau_{ABC}:=C^2_{A|BC}-C^2_{AB}-C^2_{AC}
\eeax
which is not symmetric up to the three parts $A$, $B$ and $C$).

\begin{remark}
Postulate (E4) is in consistence with the multipartite monotonic indicator 
functions of the first kind [see Eq.~(87) in Ref.~\cite{Szalay}]. From the arguments in this paper, 
the multipartite monotonic indicator functions of the second kind in 
Ref.~\cite{Szalay} is meaningless for defining MEM.
\end{remark}

\begin{remark} Hereafter, the tripartite squashed entanglement, 
	a little bit different from the one in Ref.~\cite{Yang2009ieee}, is defined by
	\bea\label{squashed-etanglement}
	E_{\rm sq}^{(3)}( \rho^{ABC}):=\frac12\inf I(A:B:C|E),
	\eea
	where 
	\beax
	I(A:B:C|E)=I(A:B|E)+I(C:AB|E),
	\eeax 
	$I(A:B|E)$ is the conditional
	mutual information, i.e., 
	\beax
	I(A:B|E)=S(AE)+S(BE)-S(ABE)-S(E),
	\eeax and
	where the infimum is taken over all extensions $\rho^{ABCE}$ of
	$\rho^{ABC}$, i.e., over all states satisfying
	$\tr_E(\rho^{ABCE})=\rho^{ABC}$. 
	In Ref.~\cite{Yang2009ieee}, the tripartite squashed entanglement, denoted by 
	$E_{\rm sq}^q$, is defined by 
	$E_{\rm sq}^{q}(\rho^{ABC}):=\inf I(A:B:C|E)$. 
	Observe that 
	\beax
	E_{\rm sq}^{(3)}\left( \rho^{ABC}\right)& =&\frac12\inf\left[
	S\left( \rho^{AE}\right) +S\left( \rho^{BE}\right) 
	+S\left( \rho^{CE}\right)\right.\nonumber\\
	&&~~~~~~~~\left.-S\left( \rho^{ABCE}\right) 
	-2S\left( \rho^E\right) \right]
	\eeax
	by definition Eq.~\eqref{squashed-etanglement},
	it is immediate that this formula
	is symmetric with respect to the subsystems $A,B,C$ though parties
	$A,B,C$ in the definition is asymmetric. Therefore we conclude that
	$E_{\rm sq}^{(3)}$ is a unified tripartite monotone.
\end{remark}

\section {Complete monogamy relation for multipartite measure}

\subsection{Complete monogamy relation for unified MEM}

Since there is no bipartite cut among the subsystems when we consider the genuine MEM, 
we thus, following the spirit of the bipartite case
proposed in~\cite{GG}, give the following definition of monogamy for
the unified tripartite measure of entanglement.

\begin{definition}\label{multimonogamy2}
Let $E^{(3)}$ be a unified tripartite entanglement measure. 
$E^{(3)}$ is said to be {completely monogamous} if for any 
$\rho^{ABC}\in\mathcal{S}^{ABC}$ that satisfies
\be\label{condofm2}
E^{(3)}(\rho^{ABC})=E^{(2)}(\rho^{AB})
\ee
we have that $E^{(2)}(\rho^{AC})=E^{(2)}(\rho^{BC})=0$.
\end{definition}

\begin{figure}
\includegraphics[width=84mm]{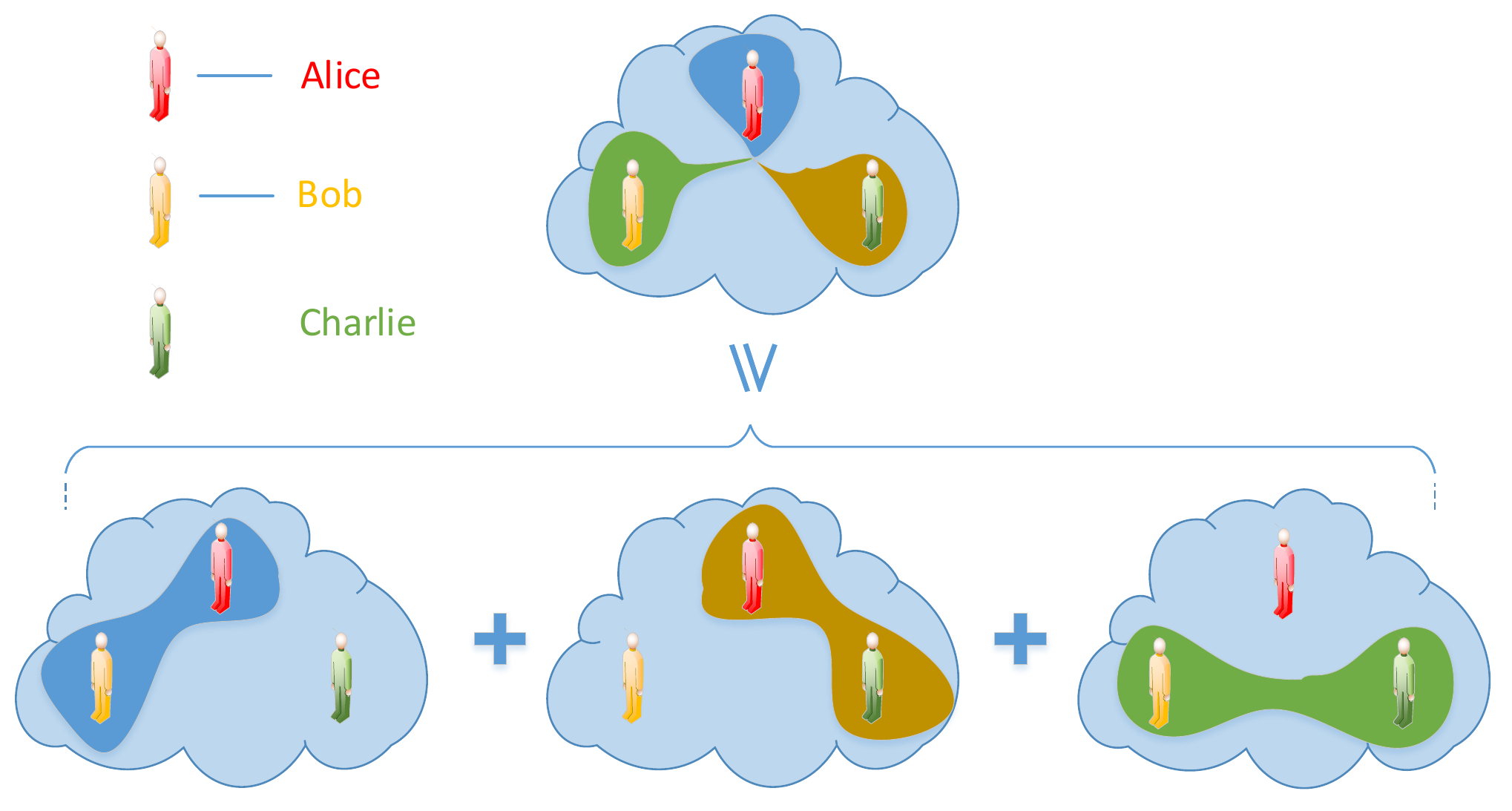}
{(a)}
\includegraphics[width=85mm]{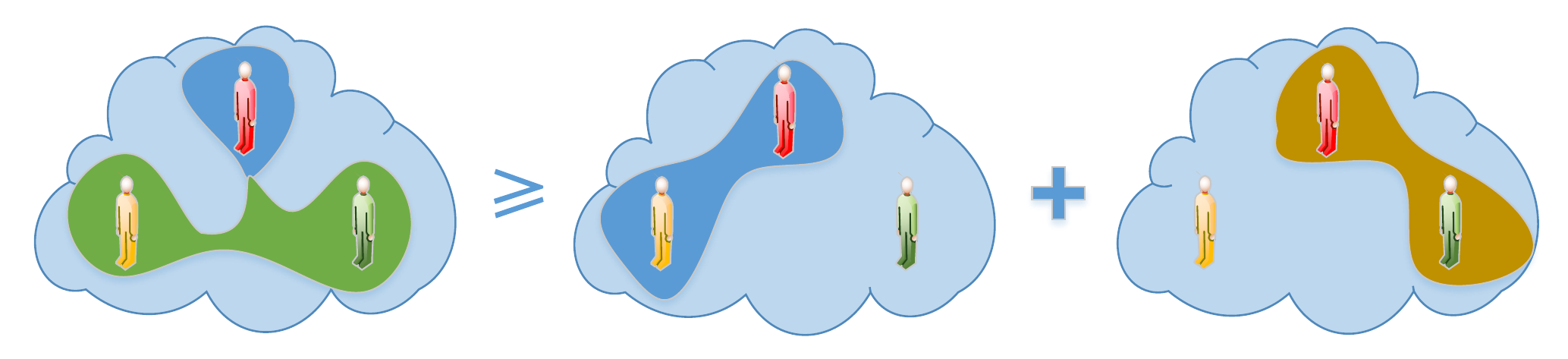}
{(b)}
\vspace{2mm}
\caption{\label{fig1}(color online). Schematic picture of 
	the monogamy relation under (a) 
	the unified tripartite entanglement measure and (b)
	the bipartite entanglement measure, respectively.}
\end{figure}

We remark here that, for tripartite measures,
the subsystem $A$ and $B$ are symmetric in the \textit{tripartite
disentangling condition}~\eqref{condofm2},
which is different from that of the bipartite disentangling 
condition~\eqref{definition-bipartite-monogamy}.
The {tripartite
disentangling condition}~\eqref{condofm2}
means that, for a given tripartite state shared by Alice, Bob, and
Charlie, if the entanglement between $A$ and $B$ reached the
``maximal amount'' which is limited by the ``total amount'' of the
entanglement contained in the state, i.e., $E^{(3)}(ABC)$, then both
part $A$ and part $B$ can not be entangled with part $C$
additionally. While the monogamy relation up to bipartite measures
is not ``complete'' (we can call it ``partial monogamy relation''), 
Definition~\ref{multimonogamy2} 
(or Proposition~\ref{monogamypower} below) captures the
nature of the monogamy law of entanglement since 
it reflects the distribution of entanglement
thoroughly and we thus call it is \textit{completely monogamous}. 
The difference between
these two kinds of monogamy relations, i.e., Eq.~\eqref{power1} and
Eq.~\eqref{power2} (see below) [or equivalently,
Eq.~\eqref{definition-bipartite-monogamy} and Eq.~\eqref{condofm2}],
is illustrated in Fig.~\ref{fig1}.
By the proof of Theorem 1
in~\cite{GG}, the following theorem is obvious.

\begin{pro}\label{monogamypower}
Let $E^{(3)}$ be a continuous unified tripartite 
entanglement measure. Then, $E^{(3)}$ is completely monogamous
if and only if there exists
$0<\alpha<\infty$ such that
\bea\label{power2}
E^{\alpha}(\rho^{ABC})\geqslant  E^{\alpha}(\rho^{AB})
+ E^{\alpha}(\rho^{AC})+ E^{\alpha}(\rho^{BC}),
\eea
for all $\rho^{ABC}\in\mathcal{S}^{ABC}$ with fixed $\dim\mH^{ABC}=d<\infty$,
here we omitted the superscript $^{(3)}$ of $E^{(3)}$ for brevity.
\end{pro}

As the monogamy exponent $\alpha$ in Eq.~\eqref{power1} for bipartite measure, we call 
the smallest possible value for $\alpha$ satisfies Eq.~\eqref{power2} in a given dimension 
$d=\dim\mH^{ABC}$, the \textit{monogamy exponent} associated with a unified measure $E^{(3)}$,
and identify it with $\alpha(E^{(3)})$. That is,
the completely monogamous measure $E^{(3)}$ together with its monogamy exponent $\alpha(E^{(3)})$
exhibit the monogamy relation more clearly. In general, the monogamy exponent is hard to calculate.
It is worth mentioning that almost all entanglement measures 
by now are continuous~\cite{GG}.
Hence, it is clear that, to decide whether $E^{(3)}$ is completely monogamous, 
the approach in Definition~\ref{multimonogamy2}
is much easier than the one from Proposition~\ref{monogamypower}
since we only need to check the states that satisfying the tripartite disentangling
condition in~\eqref{condofm2} while all states should 
be verified in Eq.~\eqref{power2}.

Let $E_F^{(3)}$ be a unified TEM defined as Eq.~\eqref{m-partite-eofmin}.
By replacing $E_f(A|BC)$, $E_f(A|B)$ with $E_F^{(3)}$, 
$E_F^{(2)}$ in Theorem 3 in Ref.~\cite{GG} respectively, 
we can conclude that, if $E_F^{(3)}$ is completely monogamous
on pure tripartite states in $\mH^{ABC}$,
then it is also completely monogamous on tripartite mixed states acting on $\mH^{ABC}$.

The first disentangling theorem was investigated in 
Ref.~\cite{Hehuan} with respect to bipartite negativity.
Very recently, Guo and Gour showed in Ref.~\cite{GG} that, 
the disentangling theorem
is valid for any bipartite entanglement monotone on pure 
states and also valid for any bipartite 
convex-roof extended measures so far.
We present here the analogous one up to tripartite measures.
One can check, following the argument of Theorem 4 and 
Corollary 5 in Ref.~\cite{GG}, that
the Lemma~\ref{lemma1} below is valid.

\begin{lemma}\label{lemma1}
	 Let $E^{(3)}$ be a unified 
	tripartite entanglement monotone, and let $\rho^{ABC}$ be a pure tripartite
	state satisfying the disentangling condition~(\ref{condofm2}). Then,
	\be
	E^{(2)}\left(\rho^{AB}\right)=E_F^{(2)}(\rho^{AB})=E_a^{(2)}(\rho^{AB})\;,
	\ee
	where $E_F^{(2)}$ is defined as in~\eqref{m-partite-eofmin}, 
	and $E_a^{(2)}$, is also defined as in~\eqref{m-partite-eofmin}
	but with a maximum replacing the minimum.
\end{lemma}

By Lemma~\ref{lemma1} we have the following result that
characterizes the form of the states that satisfying
the tripartite disentangling condition in detail.

\begin{theorem}\label{monogamyof-m-EoF}
Let $E^{(3)}$ be a unified TEM for which $h^{(2)}$, 
induced from $h^{(3)}$ as defined in~\eqref{m-h},  
is strictly concave. Then,
if  $\rho^{ABC}$ is a tripartite state and
$E_F^{(3)}(\rho^{ABC})=E_F^{(2)}(\rho^{AB})$,
then
\be
\rho^{ABC}=\sum_{x}p_x|\psi_x\ra\la\psi_x|^{ABC},
\ee
where $\{p_x\}$ is some probability distribution, 
and each pure state $|\psi_x\ra^{ABC}$ admits the form 
\be\label{product}
|\psi\ra^{ABC}=|\phi\ra^{AB}|\eta\ra^{C}\,.
\ee
\end{theorem}

\begin{proof}
By  Lemma~\ref{lemma1}, we can derive that
if $\rho^{ABC}$ be a pure tripartite
	state satisfying the disentangling condition~(\ref{condofm2}), then
	\bex
	E^{(2)}(\rho^{AB})=E_F^{(2)}(\rho^{AB})=E_a^{(2)}(\rho^{AB}),
	\eex
	where $E_F^{(2)}$ is defined as in~\eqref{m-partite-eofmin}, 
	and $E_a^{(2)}$, is also defined as in~\eqref{m-partite-eofmin}
	but with a maximum replacing the minimum.
	Let $\rho^{AB}=\sum_{j=1}^{n}p_j|\psi_j\lr\psi_j|^{AB}$
	be an arbitrary pure state decomposition of $\rho^{AB}$ with 
	$n=\rank(\rho^{AB})$. Then,
	\be\nonumber
	E^{(2)}(  \rho^{AB})  
	\leqslant E_F^{(2)}(  \rho^{AB})  
	=\sum_{j=1}^{n}p_jE^{(2)}(  |\psi_j\lr\psi_j|^{AB}).
	\ee
	On the other hand,
	\be\nonumber
	E_F^{(2)}(  \rho^{AB})  
	\leqslant E^{(3)}(   |\psi\lr\psi|^{ABC})  =h^{(3)}( \rho^A\otimes\rho^B\otimes\rho^C).
	\ee
	Therefore, denoting by $\rho_j^{A,B}:=\tr_{B,A}|\psi_j\lr\psi_j|^{AB}$,
	we conclude that if Eq.~\eqref{condofm2} holds then we must have
	\beax
	\sum_{j=1}^{n}p_jh^{(2)}(  \rho_j^A\otimes\rho_j^B)  =h^{(2)}(  \rho^A\otimes\rho^B).
	\eeax
	Given that $\rho^A=\sum_{j=1}^{n}p_j\rho_j^A$, 
	$\rho^B=\sum_{j=1}^{n}p_j\rho_j^B$, and $h^{(2)}$ is strictly concave we must have
	\be
	\rho_{j}^A=\rho^A, ~\rho_{j}^B=\rho^B, \quad j=1,...,n.
	\ee
	This leads to $|\psi\ra^{ABC}=|\psi\ra^{AB}|\psi\ra^C$.
	The case of mixed state can be easily followed.
\end{proof}

Comparing with Theorem in Ref.~\cite{GG2019}, we can see that
the strict concavity of $h^{(2)}$ for tripartite case is stronger than
that of bipartite case, which leads to that the sate satisfying the
tripartite disentangling condition just is a special case of
the one satisfying the bipartite disentangling condition.
This also indicates that the complete monogamy formula is really different from
the previous monogamy relations up to the bipartite measures.

For the case of $m$-partite case, $m\geqslant4$, we can easily derive the following 
disentangling conditions with the same spirit as that of tripartite disentangling 
condition in mind (we take $m=4$ for example):
Let $E^{(4)}$ be a unified tripartite entanglement measure. 
$E^{(4)}$ is said to be \emph{monogamous} if (i) either for any 
$\rho^{ABCD}\in\mathcal{S}^{ABCD}$ that satisfies
\be\label{condofm4}
E^{(4)}(\rho^{ABCD})=E^{(2)}(\rho^{AB})
\ee
we have that $E^{(2)}(\rho^{AB|CD})=E^{(2)}(\rho^{CD})=0$,
or (ii) for any 
$\rho^{ABCD}\in\mathcal{S}^{ABCD}$ that satisfies
\be\label{condofm5}
E^{(4)}(\rho^{ABCD})=E^{(3)}(\rho^{ABC})
\ee
we have that $E^{(2)}(\rho^{ABC|D})=0$.

The difference between the two kinds of disentangling conditions can also be revealed
by the following theorem, which is complement of the Theorem in Ref.~\cite{GG2019}.

\begin{theorem}\label{reduced-state-is-product-state}
	Let $E^{(2)}$ be an entanglement monotone for which $h^{(2)}$, as defined in Eq.~\eqref{h}, is strictly concave, and let $|\psi\ra^{ABC}$ 
	be a pure state in $\mH^{ABC}$.
	Then, 
	\beax 
	E^{(2)}(\rho^{AB})=E^{(2)}(|\psi\ra^{A|BC})~~ {\rm iff}~~ \rho^{AC}=\rho^A\otimes\rho^C,
	\eeax 
	and in turn iff 
	\beax
	|\psi\ra^{ABC}=|\psi\ra^{AB_1}|\psi\ra^{B_2C}
	\eeax 
	for some subspaces $\mH^{B_1}$
	and $\mH^{B_2}$ in $\mH^B$ up to some local unitary on part $B$, 
	where $|\psi\ra^{AB_1}\in\mH^{AB_1}$,
	$|\psi\ra^{B_1C}\in\mH^{B_1C}$;
	If $\rho^{AC}$ is separable but $\rho^{AC}\neq\rho^A\otimes\rho^C$, then $E^{(2)}(\rho^{AB})<E^{(2)}(|\psi\ra^{A|BC})$.
\end{theorem}

\begin{proof}
	Let$|\psi\ra^{ABC}$ be a pure state.
	If $\rho^{AC}=\rho^A\otimes\rho^C$,
	we assume that ${\rm rank}(\rho^A)=m$
	with spectrum decomposition
	$\rho^A=\sum_i\left( \lambda_i^A\right)^2|\psi_i\ra\la\psi_i|^A$
	and ${\rm rank}(\rho^C)=n$ with spectrum decomposition $\rho^C=\sum_j\left( \lambda_j^C\right)^2|\psi_j\ra\la\psi_j|^C$.
	It follows that
	$|\psi\ra^{ABC}$ admits the form:
	\beax
	|\psi\ra^{ABC}=\sum_{i,j}\lambda_i^A\lambda_j^C|\psi_i\ra^A|\psi_{ij}\ra^B|\psi_j\ra^C
	\eeax
	with $\la\psi_{ij}|\psi_{kl}\ra^B=\delta_{ik}\delta_{jl}$.
	Let $\mK:={\rm span}\{|\psi_{ij}\ra^B\}\subseteq\mH^B$,
	then $\mK\cong\mH^{B_1}\otimes \mH^{B_2}$
	for some subspaces $\mH^{B_1}$ and $\mH^{B_2}$.
	We thus conclude that there exists
	a unitary operator $U^B$ acting on $\mH^B$
	such that
	\beax
	U^B|\psi_{ij}\ra^{B}=|x_i\ra^{B_1}|y_j\ra^{B_2},\quad\forall~ i, j.
	\eeax
	This implies that
	\beax
	|\psi\ra^{ABC}=|\psi\ra^{AB_1}|\psi\ra^{B_2C}
	\eeax
	with $|\psi\ra^{AB_1}=\sum_i\lambda_i^A|\psi_i\ra^A|x_i\ra^{B_1}$
	and 
	$|\psi\ra^{B_2C}=\sum_j\lambda_j^C|y_j\ra^{B_2}|\psi_i\ra^C$up to local unitary operator $U^B$.
	It is now clear that $E(\rho^{AB})=E(|\psi\ra^{A|BC})$.

	Together with Theorem in~\cite{GG2019},
	we get
	\beax
	\rho^{AC}=\rho^A\otimes\rho^C\Leftrightarrow E^{(2)}(\rho^{AB})=E^{(2)}(|\psi\ra^{A|BC}).
	\eeax
	That is, if $\rho^{AC}$ is separable but $\rho^{AC}\neq\rho^A\otimes\rho^C$, then $E^{(2)}(\rho^{AB})<E^{(2)}(\rho^{A|BC})$.
	For example,
	we let 
	\beax
	|\psi\ra^{ABC}=\sum_k\lambda_k|k\>^A|k\>^B|k\>^C
	\eeax
	be a generalized GHZ state,
	then $\rho^{AC}$ is separable but $\rho^{AC}\neq\rho^A\otimes\rho^C$ and $E^{(2)}(\rho^{AB})=0<E^{(2)}(|\psi\ra^{A|BC})$.
\end{proof}

\begin{figure}
	\includegraphics[width=85mm]{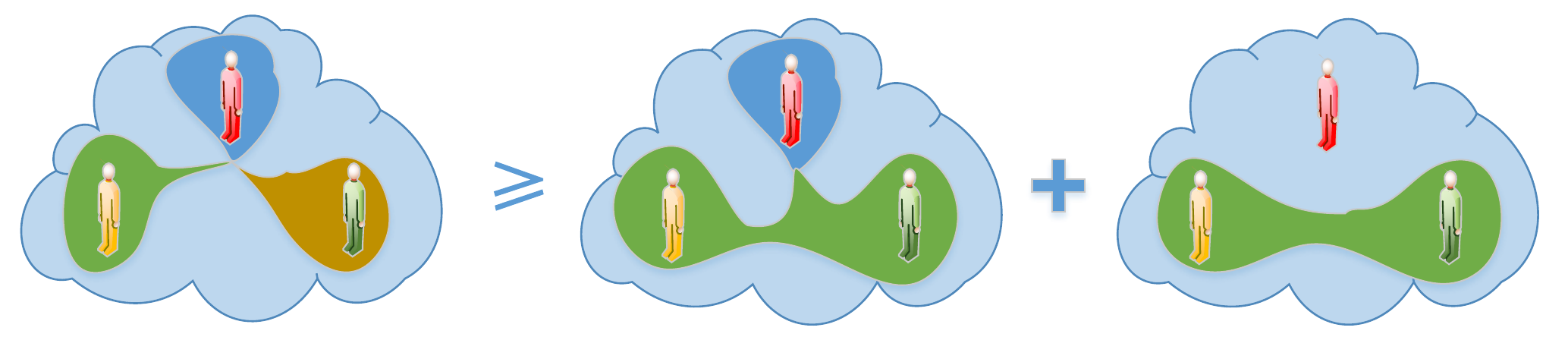}
	\caption{\label{fig}(color online). Schematic picture of the tight monogamy relation.}
\end{figure}

The bipartite squashed entanglement is shown to be
monogamous~\cite{Christandl2004jmp} with monogamy exponent is
at most 1. We prove here $E_{\rm sq}^{(3)}$ is complete monogamous.

\begin{pro}
	$E_{\rm sq}^{(3)}$ is completely monogamous, i.e.,
	\be
	E_{\rm sq}^{(3)}\left( \rho^{ABC}\right) \geqslant E_{\rm sq}\left( \rho^{AB}\right) 
	+E_{\rm sq}\left( \rho^{AC}\right) +E_{\rm sq}\left( \rho^{BC}\right) 
	\ee
	holds for any $\rho^{ABC}\in\mS^{ABC}$.
\end{pro}

\begin{proof}
	By the chain rule for the conditional mutual information with any
	state extension $\rho^{ABCE}$, it is obvious that
	\beax
	&&\frac12I(A:B:C|E)\\
	&=&\frac12I(A:B|E)+\frac12I(C:A|E)
	+\frac12I(C:B|AE)\\
	&\geqslant&
	E_{\rm sq}(\rho^{AB})+E_{\rm sq}(\rho^{AC})+E_{\rm sq}(\rho^{BC}).
	\eeax
	The proof is completed.
\end{proof}

Moreover, if there exists a optimal extension $\rho^{ABCE}$ such that
$E_{\rm sq}^{(3)}(\rho^{ABC})=\frac12I(A:B:C|E)$, then
$\rho^{ABC}$  
the tripartite disentangling condition~\eqref{condofm2} with respect 
to $E_{\rm sq}^{(3)}$ iff
$\rho^{ABEC}$ is a Markov state~\cite{HaydenJozaPetsWinter}, 
which implies that
\beax 
\rho^{ABC}=\sum_{j}q_j\rho_j^{AB}\otimes\rho_j^{C},
\eeax
where $\{q_j\}$ is a probability distribution.

\subsection{Tight complete monogamy relation for genuine MEM}

For the genuine MEM, condition (E4) exhibit the relation between
$E^{(3)}(ABC)$, $E^{(2)}(A|BC)$ and $E^{(2)}(AB)$. 
This motivates us discuss the following \textit{tight complete monogamy relation} which connects
the two different kinds of monogamy relations (i.e.,
monogamy relation up to bipartite measure and the complete one) together
(see Fig.~\ref{fig}).

\begin{definition}\label{de2}
	Let $E^{(3)}$ be a genuine MEM. We call $E^{(3)}$
	is tightly complete monogamous if
	for any state $\rho^{ABC}\in\mS^{ABC}$ that satisfying
	\bea\label{condofm3}
	E^{(3)}(\rho^{ABC})=E^{(2)}(\rho^{A|BC})
	\eea
	we have $E^{(2)}(\rho^{BC})=0$.
\end{definition}

As one may expect, we show below that, the tightly complete monogamy Eq.~\eqref{condofm3} is 
stronger than the complete monogamy relation Eq.~\eqref{condofm2} in general.

\begin{theorem}
	Let $E^{(3)}$ be a genuine multipartite entanglement monotone.
	If $E^{(3)}$ is tightly completely monogamous on pure states
	and $E_F^{(3)}$ is tightly completely monogamous,
	then $E^{(3)}$ is completely monogamous on pure states
	and $E_F^{(3)}$ is completely monogamous.
\end{theorem}

\begin{proof}
	We assume that for any $|\psi\ra^{ABC}$ that satisfies 
	$E^{(3)}(|\psi\ra^{ABC})=E^{(2)}(|\psi\ra^{A|BC})$ we have
	$E^{(2)}(\rho^{BC})=0$.
	Therefore, if 
	$E^{(2)}(\rho^{AB})=E^{(3)}(|\psi\ra^{ABC})$,
	then 
	\bea
	E^{(2)}(\rho^{AB})=E^{(2)}(|\psi\ra^{A|BC})=E^{(3)}(|\psi\ra^{ABC})
	\eea
	since $E^{(2)}(\rho^{AB})\leqslant E^{(2)}(\rho^{A|BC})\leqslant E^{(3)}(\rho^{ABC})$ holds for any state $\rho^{ABC}$.
	It follows from the assumption that
	$\rho^{BC}$ is separable. Together with Theorem~\ref{reduced-state-is-product-state}, we can conclude that
	\bea
	|\psi\ra^{ABC}=|\psi\ra^{AB}|\psi\ra^C.
	\eea
	That is $\rho^{AC}$ is a product state and thus $E(\rho^{AC})=0$.
	Namely, $E^{(3)}$ is completely monogamous for any pure states.
	We can easily check that $E_F^{(3)}$ is completely monogamous.	
\end{proof}

By Definition~\ref{de2}, the following can be easily checked.

\begin{theorem}\label{m-monogamy-relation2}
	Let $E_F^{(3)}$, defined as in Eq.~\eqref{m-partite-eofmin}, 
	be a unified TEM for which $h$,
	as defined in~\eqref{m-h}, satisfies (E4$'$) with the equality holds
	iff $\rho^{BC}=\rho^B\otimes\rho^C$.
	Then $E_F^{(3)}$ is tightly completely monogamous.
\end{theorem}

\section{Extending bipartite measures to genuine multipartite measures}

\subsection{Tripartite extension of bipartite measures}

Observing that, for pure state $|\psi\ra\in\mH^{AB}$,
\beax
&&E_f^{(2)}(|\psi\ra)=E_f(|\psi\ra)=S(\rho^A)=S(\rho^B)\\
&=&\frac12S(|\psi\ra\la\psi|\big\|\rho^A\otimes\rho^B)=\frac12S(\rho^A\otimes\rho^B)\\
&=&\frac12\left[S(\rho^A) +S(\rho^B)\right], 
\eeax
where $S(\rho\|\sigma):=\tr[\rho(\ln\rho-\ln\sigma)]$ is
the relative entropy, we thus define tripartite entanglement of formation as
\bea\label{multipartitepure4}
E^{(3)}\left( |\psi\rangle\right) &:=& \frac12\left[ S\left( |\psi\rangle\la\psi|\big\|\rho^A\otimes\rho^B\otimes\rho^C\right)\right]\nonumber\\
&=&\frac12\left[S(\rho^A)+S(\rho^B)+S(\rho^C) \right]
\eea
for pure state $|\psi\ra\in\mH^{ABC}$,
and then by the convex-roof extension, i.e.,
\bea\label{multipartitepure41}
E_f^{(3)}\left(\rho^{ABC}\right)=\min_{\{p_i,|\psi_i\ra\}}\sum_ip_iE^{(3)}(|\psi_i\ra)
\eea
for mixed state $\rho^{ABC}\in\mS^{ABC}$.
$E_f^{(3)}$ coincides with the $\alpha$-entanglement entropy defined in Ref.~\cite{Szalay}.

Let $\mathcal{P}_3^2(|\psi\ra)=\{\rho^{A}\otimes\rho^{BC},
\rho^{AB}\otimes\rho^{C},\rho^B\otimes\rho^{AC}\}$, then 
\bea\label{multipartitepure4''}
E^{(3-2)}(|\psi\ra):= \frac12\min_{\sigma\in\mP_3^2(|\psi\ra)}S\left( |\psi\ra\la\psi|\big\|\sigma\right).
\eea
For any mixed state $\rho\in\mS^{ABC}$, the entanglement of formation
associated with $E^{(3)}$ and $E^{(3-2)}$ are denoted by $E_f^{(3)}$ and $E^{(3-2)}_f$,
respectively (in order to remain consistent with the
original bipartite entanglement of formation $E_f$, we call $E_f^{(3)}$ here the tripartite EoF, 
and denote by $E_f^{(3)}$ throughout this paper. 
The notation $E_F^{(m)}$, $E_F^{(m-k)}$ with capital $F$ in the subscription 
denotes other general convex-roof extended measures).

\begin{table*}%[twocolumn]
	\caption{\label{tab:table1} Comparing of  $E^{(3)}$ and $E^{(2)}$ (or $h^{(3)}$ and $h^{(2)}$ 
		for entanglement of formation $E^{(3,2)}_F$) for $E_f^{(3)}$, tripartite concurrence $C^{(3)}$, tripartite tangle $\tau^{(3)}$, tripartite Tsallis $q$-entropy of entanglement $T_q^{(3)}$, tripartite R\'{e}nyi $\alpha$-entropy of entanglement, tripartite convex roof extended negativity $N_F^{(3)}$, tripartite negativity $N^{(3)}$, 
		tripartite squashed entanglement $E_{\rm sq}^{(3)}$, tripartite conditional entanglement 
		of mutual information $E_I^{(3)}$, tripartite relative entropy of entanglement $E_r^{(3)}$, 
		tripartite geometric measure of entanglement $E_G^{(3)}$ and the three-tangle $\tau_{ABC}$.
		 M denotes $E^{(2)}$ is monogamous, CM denotes $E^{(3)}$ is completely monogamous and 
		 TCM denotes $E^{(3)}$ is tightly completely monogamous in the following.}	
	\begin{ruledtabular}
		\begin{tabular}{cccccccc}
			$E^{(3)}$              &$E^{(3)}$ or $h^{(3)}(\rho^A\otimes\rho^B\otimes\rho^C)$  & $E^{(2)}$ or $h^{(2)}(\rho^A\otimes\rho^B)$   &E3  & E4 &M& CM & TCM      \\ \colrule
			$E_f^{(3)}$            &$\frac12S(\rho^A\otimes\rho^B\otimes\rho^C)$ &$\frac12S(\rho^A\otimes\rho^B)$ &$\checkmark$&$\checkmark$&$\checkmark$~\cite{GG2019} &$\checkmark$&$\checkmark$\\
			$C^{(3)}$ &$[3- {\rm Tr}\left( \rho^A\right) ^2-{\rm Tr}\left( \rho^B\right) ^2-\tr\left( \rho^C\right) ^2]^{\frac12}$  &$[2- {\rm Tr}\left( \rho^A\right) ^2-{\rm Tr}\left( \rho^B\right) ^2]^{\frac12}$  &$\checkmark$&$\checkmark$&$\checkmark$~\cite{GG2019} &$\checkmark$ &$\checkmark$ \\
			$\tau^{(3)}$           &$3- {\rm Tr}\left( \rho^A\right) ^2-{\rm Tr}\left( \rho^B\right) ^2-\tr\left( \rho^C\right) ^2$&$2- {\rm Tr}\left( \rho^A\right) ^2-{\rm Tr}\left( \rho^B\right) ^2$&$\checkmark$&$\checkmark$&$\checkmark$~\cite{GG2019} &$\checkmark$&$\checkmark$ \\
			$T^{(3)}_q$      &$\frac12[T_q(\rho^A)+T_q(\rho^B)+T_q(\rho^C)]$ &$\frac12[T_q(\rho^A)+T_q(\rho^B)]$&$\checkmark$&$\checkmark$&$\checkmark$~\cite{GG2019} &$\checkmark$&$\checkmark$\\
			$R^{(3)}_\alpha$       &$\frac12R_\alpha(\rho^A\otimes\rho^B\otimes\rho^C)$ &$\frac12R_\alpha(\rho^A\otimes\rho^B)$ &$\checkmark$&$\times$&$\checkmark$~\cite{GG2019} &$\checkmark$&$\times$ \\
			$N_F^{(3)}$            &$\tr^2 \sqrt{\rho^A}+\tr^2 \sqrt{\rho^B} +\tr^2 \sqrt{\rho^C} -3$                      &$\tr^2 \sqrt{\rho^A}+\tr^2 \sqrt{\rho^B}-2$            &$\checkmark$&$\times$&$\checkmark$~\cite{GG2019}  &$\checkmark$&$\times$\\ 
			$N^{(3)}$              &$\|\rho^{T_a}\|_{\tr}+\|\rho^{T_b}\|_{\tr}+\|\rho^{T_c}\|_{\tr}-3$                      &$\|\rho^{T_a}\|_{\tr}+\|\rho^{T_b}\|_{\tr}-2$            &$\checkmark$&$\times$&? &$\checkmark$&$\times$ \\  
			$E^{(3)}_{\rm sq}$~\cite{Yang2009ieee}     &$\frac12\inf I(A:B:C|E)$                                                               &$\frac12\inf I(A:B|E)$                                 &$\checkmark$&$\checkmark$&$\checkmark$~\cite{Koashi}  &$\checkmark$&?\\ 
			$E_I^{(3)}$~\cite{Yang2008prl}            &$\frac12\inf[I(AA':BB':CC')-I(A':B':C')]$                                              &$\frac12\inf[I(AA':BB')-I(A':B')]$                     &$\checkmark$&$\checkmark$&?  &?&?\\ 
			$E_r^{(3)}$~\cite{Vedral98}            &$\inf_\sigma S(\rho^{ABC}\big\|\sigma_{sep}^{ABC})$                                    &$\inf_\sigma S(\rho^{AB}\big\|\sigma_{sep}^{AB})$            &$\checkmark$&?  &?&?&?\\ 
			$E_G^{(3)}$~\cite{Barnum}            &$1-\sup_{\phi}|\la\psi|\phi\ra^{ABC}|^2$                                               &$1-\sup_{\phi}|\la\psi|\phi\ra^{AB}|^2$            &$\checkmark$&?&? &?&?\\ 
			$\tau^{(3)}_{ABC}$~\cite{Coffman}     &$C^2_{A|BC}-C^2_{AB}-C^2_{AC}$                                                         &$\times$                                               &$\times$&$\times$& $\textendash$&$\textendash$&$\textendash$         
		\end{tabular}
	\end{ruledtabular}
\end{table*}

Note that, for $|\psi\ra\in\mH^{AB}$,
$\tau(|\psi\ra)$ and $N(|\psi\ra)$ 
can be rewritten as
\beax 
\tau(|\psi\ra)&=& 
2-{\rm Tr}( \rho^A) ^2-{\rm Tr}( \rho^B) ^2,\\
N(|\psi\ra) 
&=&\frac14(\tr^2 \sqrt{\rho^A} +\tr^2 \sqrt{\rho^B} -2).
\eeax
We thus give the following definitions for any $|\psi\ra\in\mH^{ABC}$ by
\bea
\tau^{(3)}(|\psi\ra)&=&3- {\rm Tr}\left( \rho^A\right) ^2-{\rm Tr}\left( \rho^B\right) ^2-\tr\left( \rho^C\right) ^2,~~~\\
C^{(3)}(|\psi\ra)&=&\sqrt{\tau^{(3)}(|\psi\ra)},\\
N^{(3)}(|\psi\ra)&=&\tr^2 \sqrt{\rho^A}+\tr^2 \sqrt{\rho^B} +\tr^2 \sqrt{\rho^C} -3
\eea
for pure states and define by the convex-roof extension for the mixed states 
(in order to coincide with the bipartite case, we denote by
$\tau^{(3)}$,
$C^{(3)}$ and
$N_F^{(3)}$ the convex-roof extensions, respectively):
%\begin{widetext}
\beax
\tau^{(3)}\left( \rho^{ABC}\right) &=&\min_{\{p_i,|\psi_i\ra\}}
\sum_ip_i\tau^{(3)}(|\psi_i\ra\la\psi_i|),\\
C^{(3)}\left( \rho^{ABC}\right) &=&\min_{\{p_i,|\psi_i\ra\}}
\sum_ip_iC^{(3)}(|\psi_i\ra\la\psi_i|),\\
N_F^{(3)}(\rho^{ABC})&=&\min_{\{p_i,|\psi_i\ra\}}
\sum_ip_iN^{(3)}\left( |\psi_i\ra\la\psi_i|\right) ,
\eeax
%\end{widetext}
where the minimum is taken over all pure-state
decomposition $\{p_i,|\psi_i\ra\}$ of $\rho^{ABC}$.
Observe that 
\beax 
N^{(3)}(|\psi\ra)=\|\rho^{T_a}\|_{\tr}+\|\rho^{T_b}\|_{\tr}+\|\rho^{T_c}\|_{\tr}-3
\eeax
for pure state $\rho=|\psi\ra\la\psi|\in\mS^{ABC}$,
we define
\be
N^{(3)}(\rho)=\|\rho^{T_a}\|_{\tr}+\|\rho^{T_b}\|_{\tr}+\|\rho^{T_c}\|_{\tr}-3
\ee
for mixed states $\rho\in\mS^{ABC}$.
By definition, all these tripartite measures are unified (see Table~\ref{tab:table1}).
It is worth mentioning here that
$E^{(3)}$ is not unique in general for a given
$E^{(2)}$ for bipartite states.
E.g., we also can define
\be 
\tau'^{(3)}(|\psi\ra^{ABC})=2\left[  1-\sqrt{ {\rm Tr}\left( \rho^A\right) ^2}\sqrt{{\rm Tr}\left( \rho^B\right)^2}\sqrt{\tr\left( \rho^C\right)^2}\right]
\ee 
for tripartite system. $\tau'^{(3)}$ does not obey (E4):
It is easy to see that, the two-qubit state $\sigma^{BC}$
with spectra $\{87/128, 37/128, 1/32,  0\}$ as Eq.~\eqref{eq:qubitmarginal} leads
to
$\tr(\sigma^B )^2\tr( \sigma^C)^2<\tr(\sigma^{BC} )^2$ 
(the existing of such state is guaranteed by result in~\cite{Bravyi2004qic}, 
also see Eq.~\eqref{eq:qubitmarginal} below).

Since the Tsallis $q$-entropy is subadditive iff $q>1$, i.e.,
\beax
T_q(\rho^{AB})\leqslant T_q(\rho^A)+T_q(\rho^B),~ q>1, \rho^{A,B}=\tr_{B,A}\rho^{AB},
\eeax
where
\beax
T_q(\rho):=(1-q)^{-1}[\tr(\rho^q)-1]
\eeax
is the Tsallis $q$-entropy, but not additive [i.e.,
$T_q(\rho\otimes\sigma)\neq T_q(\rho)+T_q(\sigma)$ in general] in general~\cite{Raggio},
we can define tripartite Tsallis $q$-entropy of entanglement by	
\bea
T^{(3)}_q(|\psi\ra):=\frac12\left[T_q(\rho^A)+T_q(\rho^B)+T_q(\rho^C) \right],~q>1
\eea
for pure state $|\psi\ra\in\mH^{ABC}$, 
and then define by the convex-roof extension for mixed states.
The R\'{e}nyi entropy is additive~\cite{Beck},
i.e.,
\beax
R_{\alpha}(\rho\otimes\sigma)=R_{\alpha}(\rho)+R_{\alpha}(\sigma),
\eeax
we thus define tripartite R\'{e}nyi $\alpha$-entropy of entanglement by
\bea
R^{(3)}_{\alpha}(|\psi\ra):=\frac12R_{\alpha}(\rho^A\otimes\rho^B\otimes\rho^C),~0<\alpha<1
\eea
for pure state and by the convex-roof extension for mixed states,
where
\beax
R_{\alpha}(\rho):=(1-\alpha)^{-1}\ln(\tr\rho^\alpha)
\eeax
is the R\'{e}nyi $\alpha$-entropy.

\subsection{Monogamy of these extended measures}

Notice in particular that, if $E^{(3)}_F$ is a TEM
defined as in Eqs.\eqref{m-partite-eofmin} and~\eqref{m-h}, then item
(E4) is equivalent to\vspace{1mm}\\
\noindent{\bf (E4$'$)}:~$h(\rho^A\otimes\rho^B\otimes\rho^C)\geqslant
h(\rho^A\otimes\rho^{BC})$, $\forall~|\psi\ra\in\mH^{ABC}$.\vspace{1mm}\\
We can show that $E_{f}^{(3)}$, $\tau^{(3)}$ and
$C^{(3)}$ satisfy (E4$'$), and furthermore, the theorem below is true.

\begin{theorem}\label{monogamous-examples}
$E_{f}^{(3)}$, $\tau^{(3)}$,
$C^{(3)}$, $T_q^{(3)}$, $R_\alpha^{(3)}$,
$N_F^{(3)}$ and $N^{(3)}$ are completely monogamous TEMs. $E_{f}^{(3)}$, $\tau^{(3)}$,
$C^{(3)}$, and $T_q^{(3)}$ are genuine TEMs while
$R_\alpha^{(3)}$, $N_F^{(3)}$ and $N^{(3)}$ are unified TEMs but not genuine TEMS. 
\end{theorem}

\begin{proof}
	The unification condition for 
	all these quantities are clear from
	definition.
	The complete monogamy of $E_{f}^{(3)}$, $\tau^{(3)}$,
	$C^{(3)}$, $T_q^{(3)}$, $R_\alpha^{(3)}$ and
	$N_F^{(3)}$ are clear by Theorem~\ref{monogamyof-m-EoF}.
	For any $\rho^{ABC}\in\mS^{ABC}$,
	if $N^{(3)}(\rho^{ABC})=N^{(2)}(\rho^{AB})$,
	i.e.,
	$\|\rho_{ABC}^{T_a}\|_{\tr}+\|\rho_{ABC}^{T_b}\|_{\tr}+\|\rho_{ABC}^{T_c}\|_{\tr}-3=	\|\rho_{AB}^{T_a}\|_{\tr}+\|\rho_{AB}^{T_b}\|_{\tr}-2$, then
	%\bea
	$\|\rho_{ABC}^{T_c}\|_{\tr}=1$ which
	implies that $\rho^{A|BC}$ is a PPT state, and therefore
	$\rho^{AC}$ and $\rho^{BC}$ are PPT states.
	For any $E^{(3)}\in\{E_{f}^{(3)}$, $\tau^{(3)}$,
	$C^{(3)}$, $T_q^{(3)}$, $R_\alpha^{(3)}$, $N_F^{(3)}\}$ 
	and any pure state $|\psi\ra^{ABC}\in\mH^{ABC}$,
	we have
	\beax
	&&E^{(3)}( |\psi\ra^{ABC})\\
	&=&\frac12[ E^{(2)}( |\psi\ra^{A|BC}) +E^{(2)}( |\psi\ra^{AB|C})
	+E^{(2)}( |\psi\ra^{B|AC})],
	\eeax
	which indicates that $E^{(3)}$ is a TEM from the fact that $E^{(2)}$ 
	is an entanglement monotone. 
	Similarly,
	one can show that $N^{(3)}$ and $N^{(3)}$ are also TEMs.

	We now show that $E_{f}^{(3)}$, $\tau^{(3)}$,
	$C^{(3)}$ and $T_q^{(3)}$ satisfy (E4$'$).
	The cases of $E_{f}^{(3)}$ and $T_q^{(3)}$ are obvious
	since $S(\rho^{AB})\leqslant S(\rho^A)+S(\rho^B)$
	and $T_q(\rho^{AB})\leqslant T_q(\rho^A)+T_q(\rho^B)$ (note that $q>1$).
	For the case of $\tau^{(3)}$, we have
	$\tau^{(3)}(|\psi\ra^{ABC})
	\geqslant
	\tau^{(2)}(|\psi\ra^{A|BC})$
	since~\cite[Theorem 2]{Audenaerta2007jmp} 
	\beax 
	1+\tr( \rho^{BC})^2\geqslant\tr( \rho^B)^2+\tr( \rho^C)^2.
	\eeax
	Therefore the case of $C^{(3)}$ is also true.

	Recall that mixed two-qubit state $\rho^{AB}$ with spectrum
	$\lambda_1\geqslant\lambda_2\geqslant\lambda_3\geqslant\lambda_4\geqslant0$
	and marginal states $\rho^A,\rho^B$ exists if and only if the
	minimal eigenvalues $\lambda_A,\lambda_B$ of the marginal states
	satisfying the following inequalities~\cite{Bravyi2004qic}:
	\begin{eqnarray}\label{eq:qubitmarginal}
	\begin{cases}
	\min(\lambda_A,\lambda_B)\geqslant\lambda_3+\lambda_4,\\
	\lambda_A+\lambda_B\geqslant\lambda_2+\lambda_3+2\lambda_4,\\
	|\lambda_A-\lambda_B|\leqslant
	\min(\lambda_1-\lambda_3,\lambda_2-\lambda_4).
	\end{cases}
	\end{eqnarray}
	Based on this result, we can find counterexamples, which shows that
	$N_F^{(3)}$ violates (E4$'$) (then $N^{(3)}$ violates (E4$'$), either).
	Specifically, we take the following two-qubit state $\rho^{BC}$ 
	with spectrum $\{327/512$, $37/128$, $37/512$, $0\}$ and two marginal states,
	i.e., $\rho^B$ and $\rho^C$ having spectra $\{7/8$, $1/8\}$ and $\{3/4$, $1/4\}$,
	respectively. Then 
	\beax 
	1+\tr^2(\sqrt{\rho^{BC}})>\tr^2(\sqrt{\rho^B})+\tr^2(\sqrt{\rho^C}).
	\eeax
	If we take another two-qubit state $\sigma^{BC}$ such that 
	$\sigma^{BC},\sigma^B$,
	and $\sigma^C$ have spectra $\{87/128$, $37/128$, $1/32$,  $0\}$, $\{7/8$, $1/8\}$ 
	and $\{3/4$, $1/4\}$, then 
	\beax 
	1+\tr^2(\sqrt{\sigma^{BC}})<\tr^2(\sqrt{\sigma^B})+\tr^2(\sqrt{\sigma^C}).
	\eeax
	Namely, $N_F^{(3)}$ and $N^{(3)}$ violates E4$'$ for pure states.
	$R_\alpha^{(3)}$ violates E4$'$ since the Renyi $\alpha$-entropy 
	is not subadditive except for
	$\alpha=0$ or 1~\cite{Aczel}.
\end{proof}

From the proof of Theorem~\ref{monogamous-examples}, we can conclude that
if $E_{F}^{(3)}$ satisfies (E4$'$) with the equality holds
iff $\rho^{BC}=\rho^B\otimes\rho^C$
for
$|\psi\ra^{ABC}=|\phi\ra^{AB_1}|\eta\ra^{B_2C}$, 
then it is completely monogamous, but not necessarily tightly completely monogamous
as~\eqref{condofm3}.

\begin{pro}
	$E_{f}^{(3)}$, $C^{(3)}$, $
	\tau^{(3)}$ and $T^{(3)}_q$ are tightly completely monogamous
	while $R_\alpha^{(3)}$, $N_F^{(3)}$ and $N^{(3)}$ 
	are not tightly completely monogamous.
	
\end{pro}

\begin{proof}
	$R_\alpha^{(3)}$, $N_F^{(3)}$ and $N^{(3)}$ are not 
	tightly completely monogamous since they
	do not satisfy item (E4).
	Since
	$S(\rho^{BC})
	\leqslant S(\rho^{B})+S(\rho^{C})$
	holds for any pure state $|\psi\ra\in\mH^{ABC}$,
	$E_{f}^{(3)}(ABC)\geqslant E_{f}^{(2)}(A|BC)$
	for any $\rho\in\mS^{ABC}$.
	In addition, $\rho^{BC}=\rho^B\otimes\rho^C$ provided
	$E_{f}^{(3)}(|\psi\ra^{ABC})=E_{f}^{(2)}(|\psi\ra^{A|BC})$. 
	Thus $E_f^{(3)}$ is tightly completely monogamous 
	by Theorem~\ref{m-monogamy-relation2}.
	Observe that
	\beax
	\tau^{(3)}( |\psi\ra^{ABC}) 
	&=&3-\left[\tr\left(  \rho^A \right) ^2+\tr\left(  \rho^B\right) ^2 +\tr\left(  \rho^C\right) ^2\right] \\ 
	&\geqslant&2-\left[  \tr\left( \rho^A\right) ^2+\tr\left( \rho^{BC}\right) ^2 \right] \\
	&=&
	\tau^{(2)}( |\psi\ra^{A|BC})
	\eeax
	since $1+\tr( \rho^{BC})^2\geqslant\tr( \rho^B)^2+\tr( \rho^C)^2$~\cite[Theorem 2]{Audenaerta2007jmp}.
	By Proposition 4.5 
	in Ref.~\cite{Linden2013prs}, we can get
	the following result (i.e., Lemma~\ref{lemma2}, see in the Appendix for detail):
    For any bipartite state $\rho\in\mS^{AB}$, $1+\tr(\rho^2)=
		\tr(\rho^A)^2+\tr(\rho^B)^2$ if and only if
		$\rho=\rho^A\ot\rho^B$ with
		$\min\lbrace \rank( \rho^A) ,\rank( \rho^B) \rbrace =1$.
	This guarantees that 
	\beax 
	1+\tr( \rho^{BC})^2=\tr( \rho^B)^2+\tr( \rho^C)^2
	\eeax
	 if and only if $\rho^B$ or $\rho^C$ is pure.
	 For the Tsallis entropy, we have~\cite{Raggio} 
	 \bea
	 T_q(\rho\otimes\sigma)=T_q(\rho)+T_q(\sigma)
	 \eea
	 if and only if either of $\rho$, $\sigma$ is pure.
	By Theorem~\ref{m-monogamy-relation2},
	$C^{(3)}$, $\tau^{(3)}$ and $T^{(3)}_q$ are tightly completely monogamous.
\end{proof}

For $E\in\{E_{f}^{(3)}, C^{(3)},
\tau^{(3)}, T_q^{(3)}\}$, with some abuse of notations,
by Proposition~\ref{monogamypower}, Eq.~\eqref{condofm3} holds iff
\beax
E^{\alpha_1}\left( \rho^{ABC}\right)  \geqslant  E^{\alpha_1}\left( \rho^{A|BC}\right) 
+ E^{\alpha_1}\left( \rho^{BC}\right) 
\eeax
 for some ${\alpha_1}>0$.
In addition,
\beax
E^{\alpha_2}\left( \rho^{A|BC}\right) \geqslant  E^{\alpha_2}\left( \rho^{AB}\right) 
+ E^{\alpha_2}\left( \rho^{AC}\right) 
\eeax
for some ${\alpha_2}>0$ from Theorem 1 in Ref.~\cite{GG}.
Taking ${\alpha}=\max\{{\alpha_1},{\alpha_2}\}$,
we have
\beax
E^{\alpha}\left( \rho^{ABC}\right) &\geqslant& E^{\alpha}\left( \rho^{A|BC}\right) 
+E^{\alpha}\left( \rho^{BC}\right)\\
& \geqslant &E^{\alpha}\left( \rho^{AB}\right) + E^{\alpha}\left( \rho^{AC}\right) 
+E^{\alpha}\left( \rho^{BC}\right) 
\eeax
holds for these $E$.

\subsection{Additivity of the entanglement of formation}

As a byproduct of the tripartite entanglement of formation $E_f^{(3)}$, 
we discuss in this section
the additivity this measure.
Recall that, the additivity of the entanglement formation $E_f^{(2)}$ is a long
standing open problem which is conjectured to be
true~\cite{Plenio2007qic} and then disproved 
by Hastings in 2009~\cite{Hastings}.
We always expect intuitively that the measure of
entanglement should be additive in the sense of~\cite{Vollbrecht}
\bea\label{suppl-additive}
E\left( \rho^{AB}\otimes\sigma^{A'B'}\right) =E\left( \rho^{AB}\right) +E\left( \sigma^{A'B'}\right) ,
\eea
where $E(\rho^{AB}\otimes\sigma^{A'B'})
:= E(\rho^{AB}\otimes\sigma^{A'B'})$ up to the partition $AA'|BB'$.
Eq.~\eqref{suppl-additive} means that, 
from the resource-based point of view, sharing two particles from the
same preparing device is exactly ``twice as useful'' to Alice
and Bob as having just one.
By now, we know that
the squashed entanglement~\cite{Christandl2004jmp} and the
conditional entanglement of mutual information~\cite{Yang2008prl}
are additive. 
Although EoF is not additive for all states, construction of additive states for
EoF is highly expected~\cite{Zhaolijun2019pra}.
In what follows, we present a new class of states such that $E_{f}^{(2)}$ 
is additive (and thus for such class of states, we have $E_f^{(2)}=E_c$~\cite{Plenio2007qic}, 
$E_c$ denotes the entanglement cost), and present, analogously, a class of states such that
$E_{f}^{(3)}$ is additive.

\begin{theorem}\label{additivity}
	(i) Let $\rho^{AB}\otimes\sigma^{A'B'}$ be a state in $\mS^{AA'BB'}$.
	If there exists a optimal ensemble $\{p_i,|\psi_i\ra^{AA'BB'}\}$ for $E_f$ [i.e., $E_f(\rho^{AB}\otimes\sigma^{A'B'} )=\sum_ip_iE(|\psi_i\ra^{AA'BB'} )  $]
	such that any pure state $|\psi_i\ra^{AA'BB'}$ is a product state, i.e.,
	$|\psi_i\ra^{AA'BB'}=|\phi_i\ra^{AB}|\varphi_i\ra^{A'B'}$ for some pure state
	$|\phi_i\ra^{AB}\in\mH^{AB}$ and $|\varphi_i\ra^{A'B'}\in\mH^{A'B'}$,
	then we have
	\begin{align}\label{additivityEoF}
	E_{f}^{(2)}\left( AB\ot A'B'\right) =E_{f}^{(2)}\left( AB\right) 
	+E_{f}^{(2)}\left( {A'B'}\right) .
	\end{align}
	(ii) Let $\rho^{ABC}\otimes\sigma^{A'B'C'}$ be a state in $\mS^{AA'BB'CC'}$.
	If there exists a optimal ensemble $\{p_i,|\psi_i\ra^{AA'BB'CC'}\}$ for $E_f^{(3)}$
	such that any pure state $|\psi_i\ra^{AA'BB'CC'}$ is a product state, i.e.,
	$|\psi_i\ra^{AA'BB'CC'}=|\phi_i\ra^{ABC}|\varphi_i\ra^{A'B'C'}$ for some pure state
	$|\phi_i\ra^{ABC}\in\mH^{ABC}$ and $|\varphi_i\ra^{A'B'C'}\in\mH^{A'B'C'}$,
	then we have
	\be
	E_{f}^{(3)}\left( {ABC}\ot {A'B'C'}\right)=E_{f}^{(3)}\left( {ABC}\right) 
	+E_{f}^{(3)}\left( {A'B'C'}\right) .
	\ee
\end{theorem}

\begin{proof}
	We only discuss the additivity of $E_f^{(3)}$,
	the case of $E_f$ can be followed analogously.

	For pure states $|\phi\ra^{ABC}\in\mH^{ABC}$ and $|\varphi\ra^{A'B'C'}\in\mH^{A'B'C'}$,
	it is clear
	since
	\beax
	&&E_{f}^{(3)}\left( |\phi\ra\la\phi|^{ABC}\otimes|\varphi\ra\la\varphi|^{A'B'C'}\right)\\
	&=&\frac12\left[ S\left(|\phi\ra\la\phi|^{ABC}\otimes
	|\varphi\ra\la\varphi|^{A'B'C'}\big\|\rho^{AA'}\otimes\rho^{BB'}\right.\right.\\
	&&\left.\left.\otimes\rho^{CC'} \right)\right] \\
	%\eeax
	%\beax 
	&=&\frac12\left[ S\left( \rho^{AA'}\right)+S\left( \rho^{BB'}\right)+S\left( \rho^{CC'}\right)\right]\\
	&=&\frac12\left[ S\left( \rho^{A}\right)+S\left( \rho^{B}\right)+S\left( \rho^{C}\right)+S\left( \sigma^{A'}\right)\right.\\
	&&\left.+S\left( \sigma^{B'}\right)+S\left( \sigma^{C'}\right)\right] \\
	&=&\frac12\left[ S\left(|\phi\ra\la\phi|^{ABC}\big\|\rho^{A}\otimes\rho^{B}\otimes\rho^{C} \right)
	+\right.\\
	&&\left.S\left(|\varphi\ra\la\varphi|^{A'B'C'}\big\|\sigma^{A'}\otimes
	\sigma^{B'}\otimes\sigma^{C'} \right)\right] \\
	&=&E_{f}^{(3)}\left( \rho^{ABC}\right)  +E_{f}^{(3)}\left( \sigma^{A'B'C'}\right),
	\eeax
	where $\rho^{xx'}=\tr_{\overline{xx'}}\left( |\phi\ra\la\phi|^{ABC}\otimes
	|\varphi\ra\la\varphi|^{A'B'C'}\right)$, $\rho^{x}=\tr_{\bar{x}}\left( |\phi\ra\la\phi|^{ABC}\right)$
	and $\sigma^{x'}=\tr_{\bar{x'}}\left( |\varphi\ra\la\varphi|^{A'B'C'}\right)$.

	Assume that both $\rho^{ABC}$ and $\sigma^{A'B'C'}$ are mixed.
	Let $\{p_i,|\psi_i\ra^{AA'BB'CC'}\}$ be the optimal ensemble that satisfying
	\beax
	E_{f}^{(3)}\left( \rho^{ABC}\otimes\sigma^{A'B'C'}\right)
	=\sum_ip_iE_{f}^{(3)}\left( |\psi_i\ra^{AA'BB'CC'}\right) .
	\eeax
	Then
	\beax
	&&\sum_ip_iE_f^{(3)}\left( |\psi_i\ra^{AA'BB'CC'}\right)\\
	&=&\sum_ip_i\left[  E_f^{(3)}\left( |\phi_i\ra^{ABC}\right) 
	+E_f^{(3)}\left( |\varphi_i\ra^{A'B'C'}\right) \right] \\
	&\geqslant& E_{f}^{(3)}\left( \rho^{ABC}\right) 
	+E_{f}^{(3)}\left( \sigma^{A'B'C'}\right) 
	\eeax
	since by assumption we have 
	\beax
	|\psi_i\ra^{AA'BB'CC'}=|\phi_i\ra^{ABC}|\varphi_i\ra^{A'B'C'}.
	\eeax

	On the other hand, let
	$\{t_i,|\phi_i\ra^{ABC}\}$
	and $\{q_j,|\varphi_j\ra^{A'B'C'}\}$be the optimal ensembles that satisfying
	\beax
	E_{f}^{(3)}\left( \rho^{ABC}\right)
	&=&\sum_it_iE_f^{(3)}\left( |\phi_i\ra^{ABC}\right),\\
	E_{f}^{(3)}\left( \sigma^{A'B'C'}\right)
	&=&\sum_jq_jE_f^{(3)}\left( |\varphi_j\ra^{A'B'C'}\right) .
	\eeax
	Writing
	$|\psi_{ij}\ra^{AA'BB'CC'}=|\phi_i\ra^{ABC}|\varphi_j\ra^{A'B'C'}$.
	It reveals that
	\beax
	&&E_{f}^{(3)}\left( \rho^{ABC}\right)+E_{f}^{(3)}\left( \sigma^{A'B'C'}\right)\\
	&=&\sum_it_iE_f^{(3)}\left( |\phi_i\ra^{ABC}\right) 
	+\sum_jq_jE_f^{(3)}\left( |\varphi_j\ra^{A'B'C'}\right) \\
	&=&\sum_{i,j}t_iq_jE_f^{(3)}\left( |\psi_{ij}\ra^{AA'BB'CC'}\right)\\
	&\geqslant& E_{f}^{(3)}\left(\rho^{ABC}\otimes\sigma^{A'B'C'} \right).
	\eeax
	The case of $\rho^{ABC}$ is pure while $\sigma^{A'B'C'}$ is mixed
	can be proved similarly.
\end{proof}

Particularly, if $\rho^{AB}$ or $\sigma^{A'B'}$ 
(resp. $\rho^{ABC}$ or $\sigma^{A'B'C'}$) is pure, then $\rho^{AB}\otimes\sigma^{A'B'}$ (resp. $\rho^{ABC}\otimes\sigma^{A'B'C'}$) is additive under $E_f^{(3)}$ (resp. $E_f^{(2)}$).
Together with the result of Hastings in Ref.~\cite{Hastings},
we conclude that, the state $\rho^{AB}\otimes\sigma^{A'B'}$ (resp. $\rho^{ABC}\otimes\sigma^{A'B'C'}$) 
that violates the additivity~\eqref{suppl-additive} definitely have a optimal pure-state decomposition
in which some pure states are not product state up to the partition $AB|A'B'$ (resp. $ABC|A'B'C'$).
Our approach is far different from that of Re.~\cite{Zhaolijun2019pra}, 
in which it is shown that, if a state with range in the entanglement-breaking space is always additive.

\section{Maximally entangled state \& the monogamy relation}

\subsection{The original definition of maximally entangled state}

The \textit{maximally entangled state} (MES), as 
a crucial quantum resource in quantum information processing tasks such as quantum teleportation~\cite{bennett1993teleporting,noh2009quantum,zhang2006experimental},
superdense coding~\cite{bennett1992communication,barreiro2008beating},
quantum computation~\cite{bennett2000quantum}
and quantum cryptography~\cite{ekert1991quantum},
has been explored considerably~\cite{peters2004maximally,ishizaka2000maximally,facchi2008maximally,
	revzen2010maximally,verstraete2001maximally,ZG2012,aulbach2010maximally,rubin2007loss,
	gerry2002nonlinear,salavrakos2017bell,du2007experimental,lougovski2005generation,
	gerry2003generation,gilbert2008use,de2013maximally,Guo2016a}.
For a bipartite system with state space $\mH^{AB}=\mH^A\otimes\mH^B$, 
$\dim\mH^A=m$, $\dim\mH^B=n$ ($m\leq n$), a pure state $|\psi\rangle^{AB}$ 
is called a maximally entangled state if and only if 
$\rho^A=\frac{1}{m}I^A$~\cite{horodecki2001distillation}, 
where $\rho^A$ is the reduced state of $\rho^{AB}=|\psi\rangle\langle\psi|^{AB}$ 
with respect to subsystem $A$. Equivalently, $|\psi\rangle^{AB}$ is an MES if and only if
\bea\label{mes1}
|\psi\rangle^{AB}=\frac{1}{\sqrt{m}}\sum_{i=1}^m|i\rangle^A|i\rangle^B,\label{1}
\eea
where $\{|i\rangle^A\}$ is an orthonormal basis
of $\mH^A$ and $\{|i\rangle^B\}$ is an orthonormal set of $\mH^B$.
An MES $|\psi\ra^{AB}$ always archives the maximal amount of entanglement for a certain
entanglement measure~\cite{ZG2012} (such as entanglement of 
formation~\cite{Bennett1996prl,Horodecki01},
and concurrence~\cite{Hill,Wootters,Rungta}).
For example, the well-known EPR states are maximally entangled pure states.

It is proved in Ref.~\cite{cavalcanti2005all} that any MES in a $d\otimes d$ system is pure. Later,
Li \emph{et al.} showed in Ref.~\cite{ZG2012}
that the maximal entanglement can also exist in mixed states for $m\otimes n$ 
systems with $n\geq 2m$ (or $m\geq 2n$).
A necessary and sufficient condition of MEMS is proposed \cite{ZG2012}:
An $m\otimes n$ ($n\geq 2m$) bipartite mixed state $\rho^{AB}$ is maximally 
entangled if and only if
\begin{eqnarray}\label{mes2}
\rho^{AB}=\sum\limits_{k=1}^rp_k|\psi_k\rangle\langle\psi_k|^{AB},~~\sum\limits_kp_k=1,~p_k\geq0,
\end{eqnarray}
where $|\psi_k\rangle^{AB}$s are maximally entangled pure states with
\begin{eqnarray}
|\psi_k\rangle^{AB}=\frac{1}{\sqrt{m}}\sum_{i=0}^{m-1}|i\rangle^A|i_k\rangle^B,
\end{eqnarray}
$\{|i\rangle^A\}$ is an orthonormal basis
of $\mH^A$ and $\{|i_k\rangle^B\}$
is an orthonormal set of $\mH^B$, satisfying $\langle i_s|j_t\rangle^B=\delta_{ij}\delta_{st}$.
Let $\mH^{B'}$ be the subspace that spanned by $\{|i_k\rangle^B:i=0,1,\dots,m-1, k=1,2,\dots,r\}$.
Then there exists a unitary operator $U^{B'}$ acting on $\mH^{B'}$ such that
\beax 
U^{B'}|i_k\ra^B=|i\ra^{B_1}|k\ra^{B_2},
\eeax 
where 
\beax
 \mH^{B_1}:={\rm span}\{|i\rangle^{B_1}:i=0,1,\dots,m-1\}
\eeax
and 
\beax 
\mH^{B_2}={\rm span}\{|k\rangle^{B_2}: k=1,2,\dots,r\}.
\eeax
That is, the MEMS $\rho^{AB}$ can be rewritten as
\bea\label{mes3}
\rho^{AB}=|\psi_+\ra\la\psi_+|^{AB_1}
\otimes\left( \sum\limits_{k=1}^rp_k|k\rangle\langle k|^{B_2}\right),
\eea
up to some local unitary on part $B$,
where 
\beax 
|\psi_+\ra^{AB_1}=\frac{1}{\sqrt{m}}\sum_{i=0}^{m-1}|i\rangle^A|i\rangle^{B_1}
\eeax
is the maximally pure state in $\mH^{AB_1}$,
$\sum_kp_k=1$, $p_k\geq0$.
The main purpose of this section is to show that $\rho^{AB}$ in Eq.~\eqref{mes2} 
[or equivalently in Eq.~\eqref{mes3}]
is not a genuine MEMS physically, there does not exist mixed MES in any bipartite systems.

\subsection{The incompatibility of MEMS and the monogamy law}

We begin with the following fact, which seems that entanglement can be freely shared.

\begin{theorem}\label{th1}
	Let $\rho^{ABC}$ be a state acting on
	$\mH^{ABC}$ with $2\dim \mH^A\leqslant \dim \mH^{B}$.
	If $\rho^{AB}={\rm Tr}_{C}\rho^{ABC}$ is a mixed state as in Eq.~\eqref{mes2},
	then
	$\rho^{AC}$ is a product state but $\rho^{BC}$ is not necessarily separable. 
\end{theorem}

\begin{proof}
	We assume with no loss of generality that
	$\rho^{AB}$ has the form as in Eq.~\eqref{mes3}
	for some subspaces $\mH^{B_1}$ and $\mH^{B_2}$ of $\mH^B$.	
	If $|\psi\ra^{ABC}$ is a state with reduced state $\rho^{AB}$,
	then it is straightforward that
	\bea\label{abc1}
	|\phi\ra^{ABC}=|\psi_+\ra^{AB_1}|\psi\ra^{B_2C}
	\eea
	with 
	\bea\label{bc}
	|\psi\ra^{B_2C}=\sum_k\sqrt{p_k}|k\ra^{B_2}|k\ra^{C},
	\eea
	where $\{|k\ra^{C}\}$ is an orthonormal set in $\mH^C$.
	It is easy to see that $\rho^{AC}=\rho^A\otimes\rho^C$ and
	$\rho^{BC}$ is entangled.

	If $\rho^{ABC}$ is a mixed state with reduced state $\rho^{AB}$ as
	assumption, we let
	\begin{eqnarray*}
		E_f(\rho^{A|BC})
		=\sum_{s=1}^lq_sE_f(|\phi_s\ra\la\phi_s|^{A|BC}).
	\end{eqnarray*}
	It follows that
	\beax
	E_f(|\phi_s\ra\la\phi_s|^{A|BC})= E_f(\rho_s^{AB})
	\eeax
	since $\ln m\geqslant E_f(A|BC)\geqslant E_f(AB)$
	for any $\rho^{ABC}$ and 
	$\sum_sq_sE_f(\rho_s^{AB})\geqslant E_f(\rho^{AB})=\ln m$,
	where $m=\dim\mH^A$, $\rho_s^{AB}=\tr_C|\phi_s\ra\la\phi_s|^{ABC}$.
	By Theorem in Ref.~\cite{GG2019}, 
	together with the assumption of $\rho^{AB}$, we have
	\beax
	|\phi_s\ra^{ABC}=|\psi_+\ra^{AB_1}|\phi_s\ra^{B_2C},
	\eeax
	where $|\psi_+\ra^{AB_1}\in\mH^A\otimes\mH^{B_1}$
	and $|\phi_s\ra^{B_2C}\in\mH^{B_2}\otimes\mH^C$.
	We now can obtain that 
	\bea\label{abc2}
	\rho^{ABC}=|\psi_+\ra\la\psi_+|^{AB_1}\otimes\rho^{B_2C},
	\eea
	where
	\bea\label{bc2}
	\rho^{B_2C}=\sum_sq_s|\phi_s\ra\la\phi_s|^{B_2C}
	\eea
	with 
	\bea\label{bc3}
	|\phi_s\ra^{B_2C}=\sum_{k=1}^r\sqrt{p_k}|e_k^{(s)}\ra^{B_2}|f_k^{(s)}\ra^C.
	\eea
	Together with the form of $\rho^{AB}$ as supposed,
	we have $|e_k^{(s)}\ra^{B_2}=|k\ra^{B_2}$.
	It is clear that $\rho^{AC}$ is a product state and $\rho^{BC}$ 
	is entangled in general in such a case.
\end{proof}

By the argument in the proof above,
we find out that,
in the state space $\mH^{ABC}$,
even $\rho^{AB}$ achieves the maximal entanglement between part $A$
and part $B$ (i.e., it is a maximally entangled state according to Ref.~\cite{ZG2012}), 
$\rho^{AC}$ and $\rho^{BC}$ are far from each other (the former one is 
a product state and the latter one can be entangled).
Furthermore, by the arguments above, if $p_k\equiv\frac1r$, $k=1$, 2, 
$\dots$, $r$, then $\rho^{BC}=\tr_A|\psi\ra\la\psi|^{ABC}$ as in 
Eq.~\eqref{bc} is also an MES according to Ref.~\cite{ZG2012}.
In such a case
\bea\label{bac1}
|\psi\ra^{BAC}=\sum_{i,k}\frac{1}{mr}\left( |i\ra^{B_1}|k\ra^{B_2}\right) 
\otimes\left( |i\ra^A|k\ra^C\right) 
\eea
is a maximally entangled pure state with respect to the cutting $B|AC$.
Let $|f_k^{(s)}\ra^C$ as in Eq.~\eqref{bc3}. If $\dim H^C\geqslant lr$, we let
\bea\label{bc-s}
|f_k^{(s)}\ra^C=|k\ra^{C_1}|s\ra^{C_2},
\eea
for some orthonormal sets $\{|k\rangle^{C_1}$: $k=1$, $\dots$, $r\}$ 
and $\{|s\rangle^{C_2}$: $s=1$ ,$2$, $\dots$, $l\}$
in $\mH^C$, 
where 
\beax 
\mH^{C_1} :={\rm span}\{|k\rangle^{C_1}: k=1, \dots, r\}
\eeax
and 
\beax 
\mH^{C_2}={\rm span}\{|s\rangle^{C_2}: s=1,2\dots,l\}. 
\eeax
Then $\rho^{B_2C}$ in Eq.~\eqref{bc2} is an MEMS according to Ref.~\cite{ZG2012}
whenever $p_k\equiv\frac1r$.
That is, if $\rho^{AB}$ is an MEMS in the sense of Ref.~\cite{ZG2012},
it is possible that $\rho^{BC}$ is also an MEMS in the sense of Ref.~\cite{ZG2012}.
In fact,
\bea\label{bac2}
\rho^{BAC}= \sum_{s=1}^l\frac1l |\phi_s\ra\la\phi_s|^{BAC} 
\eea
with 
\bea
|\phi_s\ra^{BAC}=\sum_{i,k} \frac{1}{rm}\left( |i\ra^{B_1}|k\ra^{B_2}\right) 
\otimes\left( |i\ra^A|k\ra^{C_1}|s\ra^{C_2}\right) 
\eea
is an MEMS with respect to the cutting $B|AC$ according to Ref.~\cite{ZG2012}.
Namely, $B$ can maximally entangle with $A$ and $C$ simultaneously.

However, this fact contradicts with the
monogamy law of entanglement~\cite{Coffman,Zhuxuena2014pra,Osborne,streltsov2012are,Lan16,
	Ouyongcheng2007pra2,Kim2009,kim2012limitations,Kumar,Deng,Karczewski,Bai,Oliveira2014pra,Koashi,
	Luo2016pra,Dhar,Chengshuming,Allen,Hehuan,GG2019,GG,G2019,Camalet}: 
Entanglement cannot be freely shared among many parties. In particular, if two parties $A$ and $B$ are
maximally entangled, then neither of them can share entanglement
with a third party $C$.

It is clear that for both $|\phi\ra^{ABC}$ in Eq.~\eqref{abc1} [or~\eqref{bac1}]
and $\rho^{ABC}$ in Eq.~\eqref{abc2} [or~\eqref{bac2}],
the disentangling conditions~\eqref{definition-bipartite-monogamy} and~\eqref{condofm2} 
are valid (we take $E^{(2)}=E_f^{(2)}=E_f$ and $E^{(3)}=
E_f^{(3)}$ here). In fact, we have
\begin{itemize}
	\item $E_f^{(2)}(|\psi\ra^{A|BC})= E_f^{(2)}(\rho^{AB})$ and $E_f^{(2)}(\rho\ra^{AC})=0$.
	\item $E_f^{(2)}(|\psi\ra^{B|AC})= E_f^{(2)}(\rho^{AB})+E_f^{(2)}(\rho^{BC})$.
	\item $E_f^{(2)}(|\psi\ra^{C|AB})= E_f^{(2)}(\rho^{BC})$ and $E_f^{(2)}(\rho^{AC})=0$.
	\item $E_f^{(3)}(|\psi\ra^{ABC})= E_f^{(2)}(\rho^{AB})+E_f^{(2)}(\rho^{BC})$.
	\item $E_f^{(2)}(\rho^{A|BC})= E_f^{(2)}(\rho^{AB})$ and $E_f^{(2)}(\rho\ra^{AC})=0$.
	\item $E_f^{(2)}(\rho^{B|AC})= E_f^{(2)}(\rho^{AB})+E_f^{(2)}(\rho^{BC})$.
	\item $E_f^{(2)}(\rho^{C|AB})= E_f^{(2)}(\rho^{BC})$ and $E_f^{(2)}(\rho^{AC})=0$.
	\item $E_f^{(3)}(\rho^{ABC})= E_f^{(2)}(\rho^{AB})+E_f^{(2)}(\rho^{BC})$.
\end{itemize}
That is, 
the above examples in Eq.~\eqref{bac1}
and Eq.~\eqref{bac2} indicate that, while part $B$ and part $A$ are maximally entangled
part $B$ and part $C$ can also be maximally entangled,
which is not consistent with the monogamy law of entanglement on one hand
and that they satisfy the monogamy inequality on the other hand.
So, why does this incompatible phenomenon which seems a contradiction occur?
Is the monogamy law not true, 
or is the maximally entangled state not a ``genuinely'' MES?
We show below that the maximally entangled state should be defined by its tripartite extension
with the unified entanglement measure and the monogamy of entanglement should be
characterized by the complete monogamy relation under the unified entanglement measure. That is, the multipartite entanglement and the monogamy of entanglement cannot be revealed completely by means of the bipartite measures.

\subsection{When is a mixed state an MEMS?}

We remark here that, 
both the monogamy relation with respect to bipartite measure as in 
Eq.~\eqref{definition-bipartite-monogamy} and the complete monogamy relation as in Eq.~\eqref{condofm2}
support the monogamy law of entanglement.
Although the states in Eqs.~\eqref{bac1} and~\eqref{bac2} are MEMs according to Ref.~\cite{ZG2012},
we have \bea E_f^{(3)}(\rho^{ABC})=\ln(mr)>E_f^{(2)}(\rho^{AB})=\ln m.\eea
That is, all these monogamy relations support the monogamy law of entanglement.
In other words, the monogamy relations above are compatible with the monogamy law.
We thus believe that the monogamy law is true.

On the other hand, for pure state $|\psi\ra^{AB}\in\mS^{AB}$,
if it is maximally entangled,
then any tripartite extension $|\psi\ra^{ABC}$  
(i.e., $|\psi\ra^{AB}=\tr_C|\psi\ra\la\psi|^{ABC}$) 
must admit the form of $|\psi\ra^{ABC}=|\psi\ra^{AB}|\eta\ra^C$, that is, both $A$ and $B$
cannot entangled with $C$ whenever $A$ and $B$ are maximally entangled.
And in such a case we have $E^{(2)}(|\psi\ra^{AB})=E^{(3)}(|\psi\ra^{ABC})$
for $E^{(2,3)}=E_f^{(2,3)}$.
That is, a maximal entanglement does not depend on whether a third part is added, 
it remains maximal amount of entanglement in any extended system.
Namely, for the maximally entangled state, the maximal entanglement cannot 
increase when we add a new part. 
Therefore, we give the following definition.

\begin{definition}
	Let $\rho^{AB}$ be a state in $\mS^{AB}$ with $\dim\mH^A=m\leqslant\dim\mH^B$. Then
	$\rho^{AB}$ is an MEM if and only if i) 
	\bea E_f^{(2)}(\rho^{AB})=\ln m
	\eea and ii) for any extension
	$\rho^{ABC}$ of $\rho^{AB}$ (i.e., $\rho^{AB}=\tr_C\rho^{ABC}$) we have 
	\bea E_f^{(3)}(\rho^{ABC})=E_f^{(2)}(\rho^{AB}).
	\eea
\end{definition}

By this definition, the states in Eqs.~\eqref{bac1} and~\eqref{bac2} 
are not MEMs since $E_f^{(3)}(\rho^{ABC})>E_f^{(2)}(\rho^{AB})$.
Note that this definition of MEM is compatible with the monogamy law and
makes the concept of MEM more clearly: If $\rho^{AB}$ is an MES, then by the monogamy 
of $E_f^{(3)}$, we immediately obtain that both $\rho^{AC}$ and $\rho^{BC}$ are separable.
This also indicates that the complete monogamy relation can reflects 
the monogamy law more effectively.
From Theorem~\ref{th1}, we obtain our main result:

\begin{theorem}\label{th2}
	There is no MEMS in any bipartite quantum system.
\end{theorem}

In fact, we can also show that there is no multipartite MEMS since any 
extension of MEMS would increase entanglement from the new part. 
Note that the states in Eq.~\eqref{bac1}
and Eq.~\eqref{bac2} are really maximal to some extent,
we thus propose the following definition:

\begin{definition}
	Let $\dim\mH^{ABC}$ be a tripartite state space with $\dim\mH^A=m$ and $\dim\mH^B=n\geqslant 2m$. 
	If $\rho^{AB}\in\mS^{AB}$   
	admits the form of Eq.~\eqref{mes2},
	we call it an MEMS up to part $A$.
	If $p_k\equiv\frac1r$ in Eq.~\eqref{mes2} additionally, then 
	$\rho^{AB}$ is an MEMS up to part $B$.
\end{definition}

That is, the definition of MEMS in~\cite{ZG2012} is in fact an MEMS up to part $A$ with
the assumption that $\dim\mH^B\geqslant2\dim\mH^A$.
It is clear that $\rho^{B_2C}$ in Eq.~\eqref{bc2} with $|f_k^{(s)}\ra^C$ 
as in Eq.~\eqref{bc-s} is an MEMS up to part $B_2$
whenever $p_k\equiv\frac1r$, and if $q_s\equiv\frac1l$ additionally, then
$\rho^{B_2C}$ is an MEMS up to part $C$.
We can easily check that, if $\rho^{AB}$ is an MEMS up to part $A$, then
$\rho^{A}=\frac1mI^A$, and if $\rho^{AB}$ is an MEMS up to part $B$,
then $\rho^{A}=\frac1mI^A$ and $\rho^B=\frac{1}{mr}I^{B_1B_2}$ 
for some subspace $\mH^{B_1B_2}$ of $\mH^B$.
In addition, we can conclude that the maximally
entangled state must reach the maximal entanglement for 
well-defined entanglement measure
(such as entanglement of formation, concurrence, negativity, etc.)
but there do exist states that are not genuine maximally 
entangled state (eg. the MEMS up to part $A$) also achieves the maximal amount of entanglement.
Namely, the MEMS up to one subsystem is an MES mathematically but not physically.

\section{Conclusion and discussion}

We established a ``fine grained'' framework for defining genuine MEM and proposed the 
associated complete monogamy formula. In our framework, together with the complete 
monogamy formula, we can explore mulitipartite entanglement more efficiently. We not 
only can investigate the distribution of entanglement in more detail than the previous 
monogamy relation but also can verify whether the previous bipartite measures of 
entanglement are ``good'' measures. By justification, we found that, EoF, concurrence, 
tangle, Tsallis $q$-entropy of entanglement and squashed entanglement are better than 
R\'{e}nyi $\alpha$-entropy of entanglement, negativity and relative entropy of entanglement.
In addition, we improved the definition of maximally entangled states and showed that for 
any bipartite quantum system, the only maximally entangled state is the maximally entangled 
pure state. We can conclude that the property of bipartite state is more clear when it is 
regarded as a reduced state of its extension, namely, quantum system is always not closed, 
it should be studied in a bigger picture. The most tripartite measures by now support both 
the monogamy law of entanglement and the additional protocols of multipartite entanglement 
measures and the associated complete monogamy relation we proposed. Especially, the maximally 
entangled state is highly consistent with our scenario. We believe that our results present 
new tools and new insights into investigating multipartite entanglement and other multipartite 
correlation beyond entanglement.

As a by-product, interestingly, we found a class of states that
are additive with respect to the entanglement of formation,
which would shed new light on the problem of the classical communication
capacity of the quantum channel~\cite{Plenio2007qic,Shor2004cmp}.

However, we still do not know (i) whether the tripartite conditional entanglement of mutual 
information is completely monogamous and tightly complete monogamous, (ii) whether the tripartite 
squashed entanglement is tightly completely monogamous, and (iii) whether the tripartite 
relative entropy of entanglement and the tripartite geometric measure are genuine multipartite 
entanglement measures (also see in Table~\ref{tab:table1}). We conjecture that the answers 
to theses questions are affirmative.

\begin{acknowledgements}
Y.G. is supported by the National Natural Science Foundation of
China under Grant No.~11971277, the Natural Science Foundation of
Shanxi Province under Grant No.~201701D121001, the Program for
the Outstanding Innovative Teams of Higher Learning Institutions of
Shanxi, and the Scientific Innovation Foundation of the Higher 
Education Institutions of Shanxi Province under Grant No.~2019KJ034. 
L.Z. is supported by the National Natural Science Foundation
of China under Grant No.~11971140, and also by the Zhejiang
Provincial Natural Science Foundation of China under grant 
No.~LY17A010027 and the National Natural Science Foundation of China
under Grant Nos.~11701259 and 61771174.
\end{acknowledgements}

	\appendix*
\section{Proof of Lemma~\ref{lemma2}}

By modifying the proof of Proposition 4.5 in Ref.~\cite{Linden2013prs}, we can get
the following lemma, which is necessary in order to prove $C^{(3)}$ and $
\tau^{(3)}$ are tightly monogamous. In the proof of Lemma 1, we replace the notation 
$\rho^X$ and $I^X$ by $\rho_X$ and $I_X$, respectively, for simplicity of notations.

\begin{lemma}\label{lemma2}
	For any
	bipartite state $\rho_{AB}\in\mS^{AB}$, we have
	\begin{eqnarray}
	&&1+\max\left\{\tr\left(\rho^2_A\right),
	\tr\left(\rho^2_B\right)\right\}\tr\left(\rho^2_{AB}\right)\nonumber\\
	&\geqslant&
	\tr\left(\rho^2_A\right)+\tr\left(\rho^2_B\right),
	\end{eqnarray}
	where $\rho_{A,B}=\tr_{B,A}\rho_{AB}$.
	Moreover, $1+\tr\left(\rho^2_{AB}\right)=
	\tr\left(\rho^2_A\right)+\tr\left(\rho^2_B\right)$ if and only if
	$\rho_{AB}=\rho_A\ot\rho_B$ with
	$\min\left\lbrace \rank\left( \rho_A\right) ,\rank\left( \rho_B\right) \right\rbrace =1$.
\end{lemma}

\begin{proof}
	Without loss of generality, we assume that
	$\tr\left(\rho^2_B\right)\geqslant\tr\left(\rho^2_A\right)$. Let
	$\mathrm{spec} (\rho_A)=\{x_1,x_2,\ldots\}$ and
	$\mathrm{spec} (\rho_B)=\{y_1,y_2,\ldots\}$. For any real number
	$\kappa$, we see that
	\begin{eqnarray*}
		&&\tr\left(\rho^2_A\right)+\tr\left(\rho^2_B\right) =
		\tr\left[(\rho_A\otimes I_B+ I_A\otimes\rho_B)\rho_{AB}\right]\\
		&=&\kappa +
		\tr\left[(\rho_A\otimes I_B+I_A\otimes\rho_B-\kappa I_{AB})\rho_{AB}\right]\\
		&\leqslant& \kappa+ \tr\left[(\rho_A\otimes I_B+
		I_A\otimes\rho_B-\kappa I_{AB})_+\rho_{AB}\right],
	\end{eqnarray*}
	i.e.,
	\beax
	\tr\left(\rho^2_A\right)+\tr\left(\rho^2_B\right)\leqslant
	\kappa+\tr\left(Z_\kappa\rho_{AB}\right),
	\eeax
	where $Z_k=(\rho_A\otimes I_B+I_A\otimes\rho_B-\kappa I_{AB})_+$,
	the positive part of the operator $\rho_A\otimes
	I_B+I_A\otimes\rho_B-\kappa I_{AB}$. Furthermore, we have
	\begin{eqnarray*}
		\tr\left(\rho^2_A\right)+\tr\left(\rho^2_B\right)\leqslant
		\kappa+\tr\left(Z^2_\kappa\right)\tr\left(\rho^2_{AB}\right).
	\end{eqnarray*}
	It suffices to show 
	\begin{eqnarray}
	\min\left\lbrace \kappa+\tr\left(Z^2_\kappa\right)\tr\left(\rho^2_{AB}\right)\right\rbrace \leqslant
	1+\tr\left(\rho^2_{AB}\right).
	\end{eqnarray}
	Consider now the function
	\beax
	f_\kappa(a)=\sum_j(y_j+a-\kappa)^2_+=\|(\boldsymbol{y}+a-\kappa)_+\|^2_2,
	\eeax
	where $\boldsymbol{y}+a-\kappa:=(y_1+a-\kappa,y_2+a-\kappa,\ldots)$.
	This function is convex and
	\begin{eqnarray*}
		f_\kappa(\kappa) =
		\|\boldsymbol{y}\|^2_2=\tr\left(\rho^2_B\right)\leqslant1.
	\end{eqnarray*}
	If we assume that $\kappa\geqslant
	\max_jy_j=\|\boldsymbol{y}\|=\|\rho_B\|_\infty$, then
	\begin{eqnarray*}
		f_\kappa(0) = 0.
	\end{eqnarray*}
	Hence, under this assumption, we conclude that the convex function is
	below the straight line through
	$(0,0)$, $(\kappa,\tr\left(\rho^2_B\right))$, whose equation is given by
	$y=\frac{\tr\left(\rho^2_B\right)}{\kappa}x$. It follows from the
	above discussion that
	\begin{eqnarray*}
		f_\kappa(a) \leqslant \frac{\tr\left(\rho^2_B\right)}{\kappa}a,\quad
		a\in[0,\kappa].
	\end{eqnarray*}
	Thus, if $\kappa\geqslant \|\rho_B\|_\infty$, apparently all
	$x_i\in[0,\kappa]$, then
	\begin{eqnarray*}
		\tr\left(Z^2_\kappa\right)&=&\|Z_\kappa\|^2_2=\sum^d_{i,j}(x_i+y_j-\kappa)^2_+
		= \sum_if_\kappa(x_i)\\
		&\leqslant&\sum_i \frac{\tr\left(\rho^2_B\right)}{\kappa}x_i =
		\frac1{\kappa}\tr\left(\rho^2_B\right).
	\end{eqnarray*}
	Therefore, for any $\kappa\geqslant
	\max\{\|\rho_A\|_\infty,\|\rho_B\|_\infty\}$, we have
	\begin{eqnarray*}
		\tr\left(\rho^2_A\right)+\tr\left(\rho^2_B\right)\leqslant\kappa +
		\frac1{\kappa} \tr\left(\rho^2_B\right)\tr\left(\rho^2_{AB}\right).
	\end{eqnarray*}
	Next we consider the function
	\beax
	g(\kappa) = \kappa + \frac1{\kappa}
	\tr\left(\rho^2_B\right)\tr\left(\rho^2_{AB}\right),
	\eeax
	where 
	\beax
	\kappa\geqslant\max\{\|\rho_A\|_\infty,\|\rho_B\|_\infty\}:=\kappa_0
	\eeax
	It is easy to see that $g$ is strictly convex and it has a global
	minimum at
	\beax
	\kappa_{\min}:=\|\rho_B\|_2\|\rho_{AB}\|_2
	\eeax
	with a minimum value $g_{\min}:=2\kappa_{\min}$. Clearly, $g$ is
	strictly decreasing on the interval $(0,\kappa_{\min}]$ and strictly
	increasing on
	$[\kappa_{\min},1]$.\\~\\
	(i) If $\kappa_{\min}<\kappa_0$, then
	\begin{eqnarray*}
		\min\{g(\kappa):\kappa\geqslant \kappa_0\} =
		\kappa_0+\frac1{\kappa_0}\kappa^2_{\min}.
	\end{eqnarray*}
	(ii) If $\kappa_{\min}\geqslant\kappa_0$, then
	\begin{eqnarray*}
		\min\{g(\kappa):\kappa\geqslant \kappa_0\} =2\kappa_{\min}.
	\end{eqnarray*}
	In summary, we get that
	\begin{eqnarray*}
		\min\{g(\kappa):\kappa\geqslant \kappa_0\}=\begin{cases}
			\kappa_0+\frac1{\kappa_0}\kappa^2_{\min},&\text{if
			}\kappa_{\min}<\kappa_0,\\
			2\kappa_{\min},&\text{if }\kappa_{\min}\geqslant\kappa_0.
		\end{cases}
	\end{eqnarray*}
	Therefore, since $\kappa_0\leqslant1$, we finally get that
	\begin{eqnarray*}
		\tr\left(\rho^2_A\right)+\tr\left(\rho^2_B\right)&\leqslant&
		\min\{g(\kappa):\kappa\geqslant \kappa_0\}\leqslant
		1+\kappa^2_{\min}\\
		&\leqslant& 1+\tr\left(\rho^2_{AB}\right).
	\end{eqnarray*}
	If
	$\tr\left(\rho^2_A\right)+\tr\left(\rho^2_B\right)=1+\tr\left(\rho^2_{AB}\right)$,
	then
	
	\begin{eqnarray*}
		1+\tr\left(\rho^2_B\right)\tr\left(\rho^2_{AB}\right)=
		1+\tr\left(\rho^2_{AB}\right).
	\end{eqnarray*}
	Thus $\rho_B$ is pure state. Similarly, by the symmetric of A and B,
	we can also conclude that, if
	$\tr\left(\rho^2_A\right)\geqslant\tr\left(\rho^2_B\right)$, then
	\beax
	\tr\left(\rho^2_A\right)+\tr\left(\rho^2_B\right)\leqslant
	1+\tr\left(\rho^2_A\right)\tr\left(\rho^2_{AB}\right).
	\eeax
	In such a case, we see that
	\begin{eqnarray}
		1+\tr\left(\rho^2_A\right)\tr\left(\rho^2_{AB}\right)=
		1+\tr\left(\rho^2_{AB}\right)
	\end{eqnarray}
	implies $\rho_A$ is pure.
\end{proof}

%\nocite{*}


 %\appendix*

\end{document}